\documentclass[11pt]{article}

\def\ds{\displaystyle}

\usepackage{amsmath}
\usepackage{amsfonts}
\usepackage{amssymb}\usepackage{amsmath}
\usepackage{amsfonts}
\usepackage{amssymb}

\begin{document}
\pagestyle{headings}
\renewcommand{\thefootnote}{\alph{footnote}}

\title{Classical approximations of relativistic quantum physics}
\author{Glenn Eric Johnson\\Oak Hill, VA.\\E-mail: glenn.e.johnson@gmail.com}
\maketitle

{\bf Abstract:} A correspondence of classical to quantum physics studied by Schr\"{o}\-dinger and Ehrenfest applies without the necessity of technical conjecture that classical observables are associated with Hermitian Hilbert space operators. This correspondence provides appropriate nonrelativistic classical interpretations to realizations of relativistic quantum physics that are incompatible with the canonical formalism. Using this correspondence, Newtonian mechanics for a $1/r$ potential provides approximations for the dynamics of nonrelativistic classical particle states within unconstrained quantum field theory (UQFT).


\section{Introduction}

Early in the development of quantum mechanics, Schr\"{o}dinger and Ehrenfest [\ref{schrodinger},\ref{ehrenfest},\ref{messiah}] studied classical limits as approximations for the trajectories exhibited by the expected values of locations and momenta. The dominant support of functions that describe quantum states were associated with classical particles, and trajectories result from the time translations of the Hilbert space elements. This correspondence of features in a quantum mechanical description of nature with trajectories of points in a classical configuration space description satisfies our everyday experience. This correspondence also associates the spacetime geometry of the classical configuration space with the arguments of the functions that describe the elements of the Hilbert space realization of quantum mechanics. Significantly, this correspondence of features does not require that Hilbert space operators satisfy the technical conjecture of the canonical formalism [\ref{weinberg}].

The canonical formalism conjectures a further correspondence: that classical fields correspond to Hermitian Hilbert space field operators. The canonical formalism generalizes the Dirac-von Neumann formulation of nonrelativistic quantum mechanics. In the Dirac-von Neumann formulation, classical dynamic quantities correspond to Hermitian operators in a Hilbert space realization of quantum mechanics [\ref{dirac}]. The arche\-types for this correspondence are location ${\bf X}$ and momentum ${\bf P}$ operators that satisfy the Born-Heisenberg-Jordan relation $[{\bf X},{\bf P}]=i\hbar$. However, while the correspondence of classical dynamical objects with observable features of a quantum mechanical description of nature evidently must generalize to relativistic quantum physics, the canonical formalism is not necessary. Realizations of classical dynamic quantities as Hermitian Hilbert space operators are not necessary to satisfy our experience with the approximation of nature provided by classical concepts. The canonical formalism conjectures technical properties that are not necessarily consistent with the unification of relativity and quantum mechanics. Indeed, with consideration of relativity, location ${\bf X}$ is not a Hermitian operator [\ref{wigner},\ref{johnson}] and Hilbert space projections onto subsets of space are not consistent with causality [\ref{yngvason-lqp}]. Hermitian Hilbert space field operators are realized for free fields and related constructs, but the conjecture that a classical field corresponds to a Hermitian field operator is inconsistent with interaction in available constructions for relativistic quantum physics [\ref{intro}]. Realization of Hermitian Hilbert space field operators is excluded in the unconstrained quantum field theories (UQFT) that exhibit interaction. The classical correspondence studied by Schr\"{o}dinger and Ehrenfest is less technically demanding and their more direct correspondence admits UQFT realizations that exhibit interactions of interest [\ref{intro},\ref{gej05},\ref{feymns}] but do not satisfy the technical conjecture of the canonical formalism.

The correspondence of features in the classical limit of quantum mechanics studied by Schr\"{o}dinger and Ehrenfest does not require that multiplication by real fields defines a self-adjoint Hilbert space operator, nor is the canonical formalism's extrapolation of classical Hamiltonians to high energies and short distances imposed. These assertions for the technical properties of operations in Hilbert space realizations of relativistic quantum mechanics are problematic [\ref{pct},\ref{bogo},\ref{wightman-hilbert}]. A classical limit may be more limited than the general correspondence of classical dynamic quantities with Hermitian Hilbert space operators. The canonical formalism remains an applicable procedure when the additional technical properties are satisfied, notably in nonrelativistic quantum mechanics and in free field theory.

In this note, the methods of Schr\"{o}dinger and Ehrenfest are applied to classical correspondences for relativistic quantum physics. Classical correspondences are established by approximation of the evolution of regions of the dominant support of states as classical trajectories. This correspondence applies in nonrelativistic, particle-like instances. For this note, particular selections for classical particle-like states and a particular UQFT are developed. The selected states are generated from Gaussian minimum packet product states, selected because they achieve the Heisenberg bound for simultaneous knowledge of location and momentum, and because time translations of minimum packet functions are in family for the nonrelativistic limit of the UQFT Hamiltonian. A UQFT with a single Lorentz scalar field is selected for convenience in the analysis. Results suggest that the temporal evolution of UQFT states are approximated in nonrelativistic classical particle instances by Newtonian mechanics with a $-g/r$ potential. At the level of approximation achieved here, Newtonian mechanics suffices for the classical correspondences.

This note continues the study of realizable, quantum mechanical constructions [\ref{intro},\ref{gej05}] capable of describing the phenomenology developed in the canonical formalism. To achieve realizations, UQFT depart from the technical properties of a quantum field as described in established axioms for QFT: the G\aa rding-Wightman; and Wightman functional analytic axioms [\ref{pct},\ref{bogo},\ref{wight},\ref{borchers}]. The UQFT development of relativistic quantum physics eliminates the conjecture of a correspondence of Hermitian Hilbert space field operators with real, classical fields. UQFT exhibit interactions of interest in physical spacetime but due to a lack of Hermitian field operators exhibiting interaction, the canonical formalism does not apply. For UQFT, alternative associations of classical dynamics with relativistic quantum dynamics are of interest. Significantly, the methods of Schr\"{o}dinger and Ehrenfest associate nonrelativistic classical particle dynamics with features in the temporal evolution of selected UQFT states. This correspondence applies as an approximation for low energies and large distances with the high energy, short distance behavior determined by the vacuum expectation values (VEV) of the field. The UQFT VEV demonstrably satisfy the physical characteristics of relativistic quantum physics: Poincar\'{e} covariance, causality, and a Hilbert space realization of positive energy states. The change from the Wightman functional analytic axioms is to delete the canonical formalism-motivated property that multiplication by the field defines a Hermitian Hilbert space operator. Hermitian field operators result from satisfaction of the physical characteristics for every Schwartz tempered function of spacetime. Interacting, relativistic fields are realizable when the physical states, elements of a constructed rigged Hilbert space, are labeled from a more appropriate set of functions. The key consideration in UQFT is a local, Poincar\'{e} covariant scalar product for positive energy states. When there is interaction, an algebra of Hermitian field operators is not realized within the Hilbert space of a UQFT. The Haag-Kastler algebraic axioms [\ref{bogo},\ref{type3}] are also more general but that development emphasizes properties of operator algebras, and the assumed isotony property is not exhibited in UQFT.

Contrasted with the canonical formalism, the relativistic quantum dynamics of UQFT have a distinct character. The classical limit is derived from explicit VEV using associated classical trajectories rather than specified by a Hamiltonian form. In Lagrangian QFT and nonrelativistic quantum mechanics, an interaction term in the Hamiltonian is specified and determines the dynamics, while in UQFT, the interaction results from the form of the Hilbert space scalar product and not the generator of time translation. Rather than the derivation of quantum dynamics from a classical interaction, UQFT dynamics are constrained only to achieve the characteristics of relativistic physics. A UQFT does not necessarily model a single classical force, nor any classical force. Associations with classical dynamics are a test of the physical relevance of UQFT. Newtonian mechanics is imposed as an interpretation of nonrelativistic classical particle approximations to the quantum dynamics. These classical associations apply in likelihood and only for a limited family of states that associate with nonrelativistic classical particles. Ehrenfest's theorem is applied to associate an interaction Hamiltonian and nonrelativistic quantum mechanics to UQFT using the common associations with classical trajectories. Other than this association, UQFT lacks an interaction Hamiltonian.

First, a digression to establish notation.

\subsection{Notation}\label{sec-notation}

Spacetime coordinates in four dimensions are designated $x:=t,{\bf x}$ and energy-momentum vectors are $p:=E,{\bf p}$ with ${\bf p}:=p_x,p_y,p_z$. $x,p \in {\bf R}^4$ are Lorentz four-vectors and ${\bf x},{\bf p}\in {\bf R}^3$ are spatial (Euclidean) vectors. $x^2:=t^2-{\bf x}^2$, $p^2:=E^2-{\bf p}^2$ and $px:=Et-{\bf p}\cdot{\bf x}$ use the Minkowski signature, ${\bf p}\cdot{\bf x}$ is the Euclidean dot product and ${\bf x}^2:={\bf x}\cdot {\bf x}$ is the square of the Euclidean length. In this note, the units of spacetime coordinates are length, and the energy-momentum coordinates are wave numbers with units of the inverse length. Conversion of $t$ to time is then ``time'' = $t/c$. ``Momentum'' = $\hbar p$ and ``energy'' = $\hbar c E= \hbar c \sqrt{(mc/\hbar)^2+{\bf p}^2}$. Mass $m$ is in natural units, e.g., kilograms. Classical trajectories are designated $\xi_k(\lambda)$ and are spatial vectors parameterized by a temporal parameter $\lambda$ also with units of length. Ascending or descending sequences of multiple arguments are denoted\[(x)_{j,k}:= x_j,x_{j+1},\ldots x_k\]in the ascending case, $(x)_{j,k}:= x_j,x_{j-1},\ldots x_k$ otherwise and $(x)_n:=(x)_{1,n}$. Summation notation is used for generalized functions,\[\int dx\; T(x) f(x):=T(f)\]for a generalized function $T(x)$, function $f(x)\in {\cal A}$ with argument $x\in {\bf R}^4$. $\delta(p)$ denotes the product of $\delta(E)$, $\delta(p_x)$, $\delta(p_y)$ and $\delta(p_z)$. $\tilde{f}_n((p)_n)$ denotes the Fourier transform of $f_n((x)_n)$. The adopted Fourier transform is the evident multiple argument extension of\[\tilde{f}(p):= \int \frac{dx}{(2\pi)^2}\; e^{-ipx} f(x)\]and $\tilde{T}(\tilde{f}):=T(f)$. $\overline{z}$ designates the complex conjugate of a complex number $z=\Re e(z)+i\Im m(z)$.

Terms for Hilbert space operators include that an operator $A$ with domain ${\cal D}_A$ in a Hilbert space with scalar product $\langle u,v \rangle$ is {\em Hermitian} if $\langle u,Av\rangle= \langle Au,v \rangle$ for every $u,v \in {\cal D}_A$, a Hermitian operator is {\em symmetric} if ${\cal D}_A$ is dense, and a symmetric operator is {\em self-adjoint} if ${\cal D}_A={\cal D}_{A^*}$ and $Au=A^*u$ for every $u\in {\cal D}_A$. Then self-adjoint implies symmetric implies Hermitian. It is common to substitute symmetric, maximal symmetric and self-adjoint for Hermitian, symmetric and self-adjoint, respectively, and von Neumann's terms [\ref{vonN}] were Hermitian, maximal Hermitian, and hypermaximal. Hilbert space {\em projections} refer to orthogonal projections, $P=P^2=P^*$.

The development generally follows Borchers' description of scalar field QFT [\ref{borchers}] supplemented with definitions for UQFT from [\ref{intro}]. The properties of a Wightman-functional\[\underline{W}=(W_0,W_1,W_2((x)_2)\ldots)\]for terminating sequences of functions $\underline{f}=(f_0,f_1(x_1),f_2((x)_2),\ldots)\in {\cal A}$ describes a UQFT [\ref{intro},\ref{gej05}]. The component functions $W_n((x)_n)$ of $\underline{W}$ are the vacuum expectation values (VEV) of the fields that, as a consequence of Lorentz covariance, are generalized functions but not functions on spacetime [\ref{wizimirski},\ref{wight-pt}]. An $f_n((x)_n) \in {\cal A}$ is described by $f_0\in {\bf C}$ and $\tilde{f}_n((\pm \omega,{\bf p})_n)\in {\cal S}({\bf R}^{3n})$, Schwartz tempered functions [\ref{gel2}] of the momenta ${\bf p}_k$ when each energy is on either the positive or negative mass shell $\pm \omega({\bf p}_k)$.\begin{equation}\label{omega}\omega({\bf p}_k)^2 :=\left({\ds \frac{mc}{\hbar}}\right)^2+{\bf p}_k^2\end{equation}with a mass $m>0$ and the abbreviated notation $\omega_k:=\omega({\bf p}_k)$ is used when ${\bf p}_k$ is the argument of the Fourier transform of a function from ${\cal A}$.

The Hilbert spaces of interest to relativistic quantum physics are rigged (equipped) Hilbert spaces with elements labeled by function sequences from ${\cal B}\subset {\cal A}$, functions with Fourier transforms supported only on the positive energy support of the generalized functions $\tilde{W}_n((p)_n)$. For the scalar field UQFT, the support of the $\tilde{W}_n((p)_n)$ lies entirely on mass shells; $\tilde{W}_n((p)_n)\neq 0$ implies that every $p_k^2=m^2$. Then, for every sequence of $\varphi_n((x)_n)\in {\cal A}$, there is an $\underline{f}\in {\cal B}$ defined as\begin{equation}\label{B-defn}\tilde{f}_n((p)_n) = \prod_{k=1}^n (E_k+\omega_k) \tilde{\varphi}_n((p)_n)\end{equation}and $f_0=\varphi_0$ [\ref{intro}]. The physical states are elements of the rigged Hilbert spaces with elements labeled by equivalence classes of function sequences from ${\cal B}$, equivalent in the semi-norm provided by the nonnegative, sesquilinear, Wightman functional $\underline{W}$.\begin{equation}\label{norm} \| \underline{f} \|_{\cal B}:= \sqrt{\underline{W}(\underline{f}^* \, {\bf x}\, \underline{f})}\end{equation}with\[\underline{f} \,{\bf x}\, \underline{g} := (f_0g_0, \ldots, \sum_{\ell=0}^n f_{\ell}((x)_{\ell})\,g_{n-\ell}((x)_{\ell+1,n}), \dots).\]The Hilbert space results from a bijective map of equivalence classes of elements $\underline{f}\in {\cal B}$ in the semi-norm (\ref{norm}) to a dense set of elements $|\underline{f}\rangle$ in the Hilbert space. The map,\begin{equation}\label{isometry}\langle \underline{f}|\underline{g}\rangle= \underline{W}(\underline{f}^* \,{\bf x}\, \underline{g}),\end{equation}is an isometry. The $*$-map is an automorphism of ${\cal A}$ defined by\[f_n((x)_n) \mapsto f_n^*((x)_n) := \overline{f_n}((x)_{n,1}).\]The algebra ${\cal A}$ is $*$-involutive, ${\cal A}={\cal A}^*$, but ${\cal B}^*\cap {\cal B}$ consists of $(f_0,0,0\ldots)$. The Fourier transform of the $*$-mapped function is related to the Fourier transform of $f_n((x)_n)$ by\begin{equation}\label{mapf}\widetilde{f^*}_n((p)_n) = \overline{\tilde{f}_n} (-p_n, -p_{n-1}, \ldots ,-p_1).\end{equation}$\widetilde{f^*}_n((p)_n)$ designates the Fourier transform of $f_n^*((x)_n)$, distinct from the $*$-mapping of $\tilde{f}_n((p)_n)$ and the notation is unambiguous with the convention that the $*$-map is considered only for functions on spacetime. $\underline{W}$ satisfies Poincar\'{e} covariance and microcausality for sequences from ${\cal A}$. Fields are identified by the multiplication [\ref{borchers}] in the algebra of function sequences ${\cal A}$,\[\Phi(\underline{f})\underline{g} =\underline{f}\,{\bf x}\,\underline{g} \]and for free fields, this field results in the familiar Hermitian Hilbert space operator [\ref{pct},\ref{borchers}]. ${\cal A}$ includes real functions and functions of compact spacetime support but the subset ${\cal B}$ lacks both real functions and functions of bounded support although elements of ${\cal B}$ may be arbitrarily dominantly supported within bounded regions of spacetime. For the single Lorentz scalar field UQFT,\begin{equation}\label{hamil}H=\sum_{k=1}^n \omega_k\end{equation}is the Hamiltonian in the $n$ particle subspace that results from time translations, $|U(t)\underline{f}\rangle =|\underline{g}\rangle$ with\[\tilde{g}_n((p)_n) = \prod_{k=1}^n e^{-i\omega_k t} \,\tilde{f}_n((p)_n).\]$H$ would be designated a free field Hamiltonian in a canonical formalism development of QFT. $U(t)$ is the unitary homomorphism of the group of time translations that results from translation invariance of the Wightman-functional [\ref{pct}]. Eigenstates of $H$ are plane waves realized as generalized eigenfunctions but not as elements of the Hilbert space.

\subsection{Nonrelativistic classical particle states}\label{sec-defn}

The idealization of a classical particle is used to associate UQFT with nonrelativistic classical dynamics. When a state of a UQFT can be labeled by a function described by nonrelativistic classical particles, that state is a nonrelativistic classical particle state. The properties of the temporal evolution of these states determine the classical limits of the UQFT. Gaussian functions centered on classical particle trajectories are used to describe the classical limits. The Gaussian, minimum packet functions are distinguished as the most nearly classical states in the sense that the Heisenberg bound is achieved and time translation is in family for the nonrelativistic limit of the UQFT Hamiltonian. The terms used to describe a classical particle associated with a UQFT include:

\begin{enumerate}
\item A {\em classical model} is a set of $n$ twice differentiable trajectories $\xi_k(\lambda)\in {\bf R}^3$ parameterized by a real temporal parameter $\lambda$. Trajectories describe the motion of points in ${\bf R}^3$. Each trajectory has a velocity $\dot{\xi}_k(\lambda)$, the first derivative with respect to $\lambda$. For the accuracy of the approximations considered in this note, Newton's equation $F=ma$ suffices as the equation of motion to determine trajectories from initial conditions. Physically acceptable sets of classical trajectories conserve the total momentum and the Poincar\'{e} covariance of UQFT scalar products is exploited to select coordinates collocated with the classical center of mass. In such a frame, $\sum_k \xi_k(\lambda)=\sum_k \dot{\xi}(\lambda)=0$ for any $\lambda$. A set of $n$ classical trajectories $\xi_k(\lambda)$ are denoted {\em strictly bound} if there exists an $R < \infty$ with $\|\xi_k(\lambda)\|<R$ for all $\lambda\geq 0$ and any $k$ in coordinates collocated with the center of mass of the classical trajectories. 
\item A {\em classical particle} is associated with the region of dominant support of a minimum packet function $\varphi({\bf x};{\bf \xi}_k,\lambda)$ centered on a trajectory $\xi_k:=\xi_k(\lambda)$ with a momentum ${\bf q}_k:={\bf q}_k(\lambda)$ and $\hbar {\bf q}_k(\lambda)=mc\gamma_k \dot{\xi}_k(\lambda)$. The minimum packet functions are Schwartz functions [\ref{gel2}] with Fourier transforms\[\tilde{\varphi}({\bf p};{\bf \xi}_k,\lambda):=L_0(\lambda)^3\,\exp\left(-(L_0(\lambda)^2-i\frac{\lambda_c}{2} \lambda) ({\bf p}-{\bf q}_k)^2+i ({\bf p}-{\bf q}_k)\!\cdot\!\xi_k(\lambda)\right).\]These Gaussian functions $\varphi({\bf x};{\bf \xi}_k,\lambda)$ and associated trajectories $\xi_k(\lambda)$ are described in Sections \ref{nonrel} and \ref{sec-minpak}. The Compton wavelength for mass $m$ is\begin{equation}\label{compton}\lambda_c :=\frac{\hbar}{mc}.\end{equation}The momentum support of these minimum packet functions is centered on ${\bf q}_k$ with an extent characterized by a real function $L_0(\lambda)$. The length $L_0(\lambda)$ determines the momentum variance and provides a lower bound on the spatial variance of the minimum packet states. In this note, $L_0(\lambda)$ is referred to as the {\em momentum spread length}. The $n$ particle minimum packet states are labeled by function sequences\begin{equation}\label{n-minpak} \underline{s}(\lambda):=(0,\ldots,0,s_n((x)_n,\lambda),0,\ldots) \in {\cal B}\end{equation}with Fourier transforms\[\tilde{s}_n((p)_n;\lambda):= \prod_{k=1}^n (E_k+\omega_k)\; \tilde{\varphi}({\bf p}_k;{\bf \xi}_k,\lambda)\]and $\lambda$ is the temporal parameter of the trajectories $\xi_k(\lambda)$ and $L_0(\lambda)$. A normalization of the state labeled by (\ref{n-minpak}) is designated\begin{equation}\label{n-minpak-normd}|\underline{u}_s(\lambda)\rangle := K_s(\lambda)\,|\underline{s}(\lambda)\rangle\end{equation}with $K_s(\lambda)\in {\bf R}$ and $\|\underline{u}_s(\lambda)\|_{\cal B}=1$ for the norm (\ref{norm}). These $s_n((x)_n;\lambda)$ have point support in time with each $t_k=0$ [\ref{intro}]. Functions in ${\cal B}$ include generalized functions with point support in time and as a consequence, UQFT includes descriptions for physical states at particular times. Minimum packet functions describe particles when the supports are isolated and exhibit nonrelativistic relative velocities. The concentrations in the support of the functions that label UQFT states are distinguishable until they bifurcate or merge with other peaks.
\item A state $|\underline{u}_s(\lambda)\rangle$ of unit norm is called a {\em nonrelativistic classical particle} state if it is labeled by minimum packet functions centered on the classical trajectories $\xi_k(\lambda)$, the minimum packet functions describe classical particles, and time evolution is nearly equivalent to translations of the temporal parameter of the trajectories,\begin{equation}\label{idea} |U(-\lambda) \underline{u}_s(0)\rangle \approx | \underline{u}_s(\lambda) \rangle.\end{equation}This approximation associates time translation of an appropriate subset of elements of the Hilbert space with an evolution of points described by classical mechanics.
\end{enumerate}This note studies (\ref{idea}), that temporal translation is approximately the same as a translation of particles along classical trajectories for states described by nonrelativistic classical particles. Results include that the potential associated with the UQFT is a regularized, attractive $-g/r$ potential.

The negative sign in (\ref{idea}) originates in the opposite senses of translation for functions and generalized functions.\[\renewcommand{\arraystretch}{1.75} \begin{array}{rl}\langle \underline{s}(\lambda) | U(\tau) \underline{s}(0)\rangle &= \underline{W}(\underline{s}^* \,{\bf x}\, (1,\tau) \underline{s})\\
 &= {\ds \int} d(x)_{2n}\; W_{2n}((x)_{2n})\,\overline{s_n}((x)_{n,1},\lambda) s_n((t-\tau,{\bf x})_{n+1,2n},0)\\
 &= {\ds \int} d(x)_{2n}\; W_{2n}((x)_n,(t+\tau,{\bf x})_{n+1,2n})\,\overline{s_n}((x)_{n,1},\lambda) s_n((x)_{n+1,2n},0).\end{array}\]Then, (\ref{idea}) is that\[s_n((t+\tau,{\bf x})_{n+1,2n},0) \approx s_n((t,{\bf x})_{n+1,2n},\tau)\]in the norm $\|\underline{s}\|_{\cal B}$ from (\ref{norm}), for trajectories $\xi_k(\lambda)$ other than the straight line trajectories that characterize a lack of interaction, and for UQFT VEV $W_{2n}((x)_{2n})$ evaluated for functions (\ref{n-minpak}) that describe nonrelativistic classical particles. States labeled by appropriate minimum packet functions are associated with classical particles and time translation of these Hilbert space elements provide the trajectories described by classical predictions.

The approximation (\ref{idea}) is in the trace norm and implies approximation up to a phase. The Born's rule transition likelihood\begin{equation}\label{likelihood} \mbox{Trace}(P_f |\underline{u}_s(0) \rangle \langle \underline{u}_s(0) |)\approx 1\end{equation}for appropriate, normalized nonrelativistic classical particle states (\ref{n-minpak-normd}) and an orthogonal projection operator\begin{equation}\label{proj}P_f=\int d\mu_s \; |U(\lambda)\underline{s}(\lambda)\rangle \langle U(\lambda)\underline{s}(\lambda)|\end{equation}with $\lambda\geq 0$ and $\mu_s$ a measure on subsets of the parameters describing the states $|\underline{s}(\lambda)\rangle$ in (\ref{n-minpak}). Plane wave limits as well as elements of the Hilbert space are of interest. An amplitude ${\cal I}(\lambda,\tau)$ is designated such that the likelihood (\ref{likelihood}) is\begin{equation}\label{i-defn}\int d\mu_s \; {\cal I}(\lambda,0)^2:=\mbox{Trace}(P_f |\underline{u}_s(0) \rangle \langle \underline{u}_s(0) |).\end{equation}When the projection (\ref{proj}) is onto the one dimensional subspace of the element $|U(\lambda)\underline{s}(\lambda)\rangle$, the transition likelihood (\ref{likelihood}) is the square of the magnitude of the scalar product\[\mbox{Trace}(P_f |\underline{u}_s(0) \rangle \langle \underline{u}_s(0) |)= {\cal I}(\lambda,0)^2 =|\langle U(\lambda)\underline{u}_s(\lambda)| \underline{u}_s(0)\rangle|^2.\]When the projection is onto the subspace defined by an appropriate linear combination of plane wave limits, the measure $d\mu_s$ of interest is Lebesgue measure $d({\bf q})_n$ on the classical momenta that describe plane wave limits and\[{\cal I}(\lambda,0) =N_{\cal I} |\langle U(\lambda)\underline{s}(\lambda)| \underline{u}_s(0)\rangle|.\]The normalization $N_{\cal I}$ for plane waves derives from $P_f^2=P_f$ and is included in Section \ref{sec-norm}. Then generally, nonrelativistic approximations to the scalar products\begin{equation}\label{scalarprod} \langle U(\lambda)\underline{s}(\lambda)| U(\tau)\underline{s}(\tau)\rangle\end{equation} underlie evaluation of the likelihoods (\ref{likelihood}) of interest to classical limits. ${\cal I}(\lambda,\tau)$ is a function of the trajectories $\xi_k(\tau),\xi_k(\lambda)$ and their first temporal derivatives $\dot{\xi}_k(\tau),\dot{\xi}_k(\lambda)\approx\lambda_c {\bf q}_k(\lambda)$, and momentum spread lengths $L_0(\tau),L_0(\lambda)$. These scalar products are evaluated in Section \ref{sec-eval}.

\subsection{Overview}\label{sec-cllim}

In this section, the associations of classical trajectories with the temporal evolution of regions of dominant support resulting from\[|U(-\lambda) \underline{u}_s(0)\rangle \approx | \underline{u}_s(\lambda) \rangle\]for the selected UQFT are organized and introduced. 

Trajectories $\xi_k(\lambda)$ and momentum spread lengths $L_0(\lambda)$ appropriate to descriptions of nonrelativistic classical particles satisfy constraints. Association of a classical particle with a dominant region in the support of a function describing a quantum state includes the considerations that the dominant regions are sufficiently isolated to be distinguishable with significant confidence, relative velocities are nonrelativistic and that energies are sufficiently small to neglect changes in particle number. With $r_a(\lambda)$ the minimum Euclidean distance between $n$ locations $\xi_k(\lambda)$,\[r_a(\lambda) := \min_{j,k\neq j} \|\xi_k(\lambda)-\xi_j(\lambda)\|,\]the parameters of nonrelativistic classical particle states satisfy bounds developed in Section \ref{sec-minpak}.

\begin{quote}{\bf Nonrelativistic classical particle bounds}: The association of particles with UQFT states labeled by minimum packet functions (\ref{n-minpak}) applies for times $\lambda$, trajectories $\xi_k(\lambda)$ and momentum spread lengths $L_0(\lambda)$ that satisfy\begin{equation}\label{L0-ineq} \renewcommand{\arraystretch}{1.75} \begin{array}{l} \|\dot{\xi}_k(\lambda)\| \ll 1 \\
 \lambda_c \ll L_0(\lambda)\\
L_0(\lambda)^2+{\ds \frac{\lambda_c^2\lambda^2}{4L_0(\lambda)^2}} \ll r_a(\lambda)^2\end{array}\end{equation}in a coordinate frame collocated with the classical center of mass of the $n$ particles.\end{quote}
 
For spatial variances that are a significant fraction of the minimum separation $r_a(\lambda)$, the overlap of the supports of the minimum packet functions is appreciable and reliable association of classical particles with regions of support degrades. The bounds require that the closest approach $r_a$ of the classical trajectories significantly exceeds the Compton wavelength. The nonrelativistic classical particle conditions eliminate a correspondence with classical particles for the constituents of deeply bound states. In these cases, the closest approach of the constituents is near or less than the Compton wavelength.

There are several qualifications to (\ref{idea}) in addition to satisfaction of the constraints (\ref{L0-ineq}). The accuracy of (\ref{idea}) degrades with growth of $\lambda$ as errors from the nonrelativistic approximation grow, and the approximation of time translation by changes in the temporal parameter of the trajectories is limited by the use of a single classical trajectory. The duration $\lambda$ of accurate nonrelativistic approximations varies with situation. In Section \ref{sec-connct}, a bound on the useful duration of the approximation is developed. This bound grows at least as fast as the square of the momentum spread length, $L_0(\lambda)$. As a consequence, the nonrelativistic approximations are accurate over a greater duration for wave-like states than point-like states. The nonrelativistic approximations are more accurate when the momentum spreads are small.

The approximation for the evolution of the quantum state by a single classical trajectory per particle also degrades with increases in the period $\lambda$ in (\ref{idea}). Sets of initially nearby trajectories may bifurcate rather than spread about a single trajectory. Accommodation of the uncertainty in positions and velocities as initial conditions for the classical trajectories results in a spread of associated classical trajectories. In free field theory, this uncertainty is accommodated by growth in the spatial variance of the minimum packet functions with time. In the nonrelativistic approximation to free field theory, an initial variance $L_0^2 \mapsto L_0^2 +\lambda_c^2 \lambda^2/4L_0^2$. With interaction, the description of trajectories is more diverse. In the example of a loosely bound, circular classical trajectory, trajectories that are initially nearby include scattered solutions and elliptical orbit bound solutions. Then, consideration that the spatial variance of the bound states must remain small with respect to orbit radii is incompatible with the divergence of the orbits and scattered states. A large spatial variance contradicts the distinguishability of the bound state constituents in a classical description. As a consequence, the duration of the time span over which (\ref{idea}) is applicable is limited by constraints on packet spread. And finally, deviations from Newtonian mechanics are anticipated, for example, to achieve geometrodynamical gravity and a causal radiation reaction force in electrodynamics.

With recognition of these limitations, (\ref{idea}) provides a correspondence of UQFT with trajectories from Newton's mechanics. The development here uses a common $L_0(\lambda)$ for all arguments and one generalization would be an $L_{0,k}(\lambda)$ labeled in common with a corresponding trajectory $\xi_k(\lambda)$. This generalization is not necessary to the results of this note.

Momentum spread lengths $L_0(\lambda)$ play a fundamental role in the correspondence of classical trajectories with UQFT states. Momentum spread lengths $L_0(\lambda)$ provide the freedom to associate appropriate classical limits with UQFT. A nonrelativistic classical particle description (\ref{L0-ineq}) alone is not sufficient to associate a UQFT with a classical force. A dynamical model for $L_0(\lambda)$ is selected to result in the classical particle trajectories observed in nature. Assumptions that associate UQFT with appropriate classical limits are identified in Section \ref{sec-corrsp}. The form\begin{equation}\label{L0-form}L_0(\lambda):=L_0(T(\lambda),V(\lambda))\end{equation}associates classical limits of UQFT with Newton's equations. Then, $L_0(\lambda)$ is regarded as an abbreviated notation for $L_0(T,V)$ evaluated at time $\lambda$ with $T(\lambda)$ the classical kinetic energy defined below in (\ref{it-defn}) and $V(\lambda)$ the classical potential energy defined below in (\ref{potnl}). $T(\lambda)$ and $V(\lambda)$ derive from the classical trajectories $\xi_k(\lambda)$.

The momentum spread length $L_0(\lambda)$ and the duration of the interval between observations $\lambda$ are used to organize consideration of (\ref{idea}) into domains. For $n$ particle states (\ref{n-minpak}), the scalar products (\ref{scalarprod}) are decomposed below in (\ref{expand}) of Section \ref{sec-vev} into terms with distinct cluster properties. Each term in a scalar product is designated as either a {\em forward contribution} that consists entirely of two-point function factors, a {\em partially factored contribution} that includes at least two connected function factors with at least one factor having more than two arguments, or a {\em connected contribution} that consists of a single connected function. The sum of the forward VEV contributions is designated $F(\lambda,\tau)$ and the connected VEV contribution is designated $c_{2n}C(\lambda,\tau)$ with coefficients $c_{2n}$ that determine interaction strength. In Section \ref{sec-eval} it will be demonstrated that for sufficiently large $L_0(\lambda)$, the forward VEV contribution dominates the connected and partially factored contributions, $F(\lambda,\lambda)\gg c_{2n}C(\lambda,\lambda)$. Conversely, for sufficiently small $L_0(\lambda)$, the connected VEV contribution dominates the forward and partially factored contributions, $F(\lambda,\lambda)\ll c_{2n}C(\lambda,\lambda)$. Small $L_0(\lambda)$ correspond to point-like states and large $L_0(\lambda)$ to plane wave-like states. The partially factored contributions are implicit in this notation.

Results for each domain are briefly summarized here and are developed in the referenced sections. For the summary, the forward VEV contributions $F(\lambda,0)$ to the scalar products (\ref{scalarprod}) are considered negligible for trajectories that deviate significantly from straight lines. The domains considered are:

\begin{enumerate}
\item \label{item-1} The nonrelativistic classical particle approximation for the time evolution of states (\ref{idea}) applies over short intervals. Asymptotically for small $\lambda$, the approximation (\ref{idea}) is a property of the minimum packet functions (\ref{n-minpak}) for trajectories that are solutions to Newton's equation for $1/r$ potentials. Solutions to Newton's equation for $1/r$ potentials are associated with a scalar field UQFT independently of the VEV or potential strength $g$. This result is developed in Section \ref{sec-minpak} as Lemma 1.

\item \label{item-2} Over intervals $(0,\lambda)$ with sufficiently small $L_0(\lambda)$ and when $F(\lambda,0)$ is negligible, the likelihoods (\ref{likelihood}) are nearly independent of the interaction strengths $c_{2n}$.\[\langle U(\lambda)\underline{u}_s(\lambda)|\underline{u}_s(0)\rangle \approx \frac{C(\lambda,0)}{C(\lambda,\lambda)^\frac{1}{2} C(0,0)^\frac{1}{2}}\]For small $L_0(\lambda)$, nonrelativistic approximations degrade most rapidly and the duration of the interval available to analysis using the methods of this study is therefore limited. For the example of two particles in the circular orbits of a $-g/r$ potential, the likelihood (\ref{likelihood}) is optimized for an interaction strength $g$ proportional to $(c_4)^\frac{2}{3}$ when the momentum spread length $L_0(\lambda)$ is appropriately selected. In this example of two particles in circular orbit, quantum corrections to a $-g/r$ potential appear for large separations $r$. These particle-like cases are developed in Section \ref{sec-twop}. 

\item \label{item-3} When an initial, sufficiently small $L_0(0)$ transitions to a sufficiently large $L_0(\lambda)$, accuracy of the temporal approximation persists and\[\langle U(\lambda)\underline{u}_s(\lambda)|\underline{u}_s(0)\rangle \approx \frac{\sqrt{c_{2n}}\;C(\lambda,0)}{F(\lambda,\lambda)^\frac{1}{2} C(0,0)^\frac{1}{2}}.\]In these instances, any nonrelativistic classical particle trajectories that satisfy Newton's equations is a local extremum of the approximation (\ref{idea}). The optimization includes determination of the momentum spread lengths $L_0(\lambda)$ to associate the UQFT with appropriate classical limits. These particle-like to wave-like transition cases are developed in Section \ref{sec-newteq}.

\item \label{item-4} The correspondence (\ref{idea}) of UQFT states with classical particle descriptions does not apply to plane wave scattering amplitudes. Particle-like cases with initially large $L_0(0)$ transitioning to large $L_0(\lambda)$ result in negligible interaction. To satisfy the nonrelativistic classical particle bounds, the trajectories must remain many times $L_0$ apart and then the plane wave limit places the trajectories at too great a separation to interact significantly, and the quantum corrections of item \ref{item-2} obscure the associations of states with classical trajectories at great ranges. Plane wave scattering amplitudes result from overlapping wave-like states. For $L_0(0)=L_0(\lambda)$ growing without bound, $F(\lambda,\lambda)$ is the only significant contribution to state norms and $C(\lambda,0)$ is the only significant contribution to non-forward scattering.\[\langle U(\lambda)\underline{u}_s(\lambda)|\underline{u}_s(0)\rangle \approx \frac{c_{2n}C(\lambda,0)}{F(\lambda,\lambda)^\frac{1}{2} F(0,0)^\frac{1}{2}}.\]The non-forward scattering amplitudes for the single Lorentz scalar field UQFT are\[\langle (q_1,\ldots q_n)^{\mathit{out}}|(q_{n+1},\ldots q_{n+m})^{\mathit{in}} \rangle = c_{n+m}\;\delta(q_1 \ldots\!+\!q_n \!-\!q_{n+1}\ldots \!-\!q_{n+m})\]from [\ref{intro},\ref{gej05},\ref{feymns}]. These scattering amplitudes are proportional with a factor $i$ to the first contributing order of a Feynman series [\ref{weinberg}] for a ${\cal P}(\Phi)_4$ interaction Hamiltonian density\[H_{int}(x)=\sum a_k :\!\Phi(x)^k\!:\]with $k\geq 4$, $a_k=c_k\,(2\pi)^{2k-4}/k!$, and $c_k$ are the interaction strength coefficients from the VEV in (\ref{vev-connctd}) below. The nonrelativistic limit of the corresponding differential cross section $d\sigma/d\Omega$ is not the Mott cross section for particles interacting with a $1/r$ potential despite the association of the nonrelativistic classical particle trajectories with $1/r$ potentials. There are scalar field UQFT with elastic cross sections that have nearly (regularized) $1/r$ equivalent plane wave scattering potentials in first Born approximation [\ref{feymns}], but evaluation of the scalar products (\ref{scalarprod}) in these cases is beyond the scope of this study. The association of the selected scalar UQFT with the trajectories of a $1/r$ potential is natural in the particle-like instances but not in this wave-like instance.
\end{enumerate}

A plausible bound on the scalar products (\ref{scalarprod}) joins the three particle-like domains \ref{item-1}-\ref{item-3}. Satisfaction of the bound provides that the trajectories of a $-g/r$ potential with a strength $g$ determined by $c_4$ are the trajectories of greatest likelihood.

\begin{quote}{\bf Coplanar propagation}: The states of a UQFT satisfy the {\em coplanar propagation bound} if the magnitudes of scalar products of states\[|\underline{\psi}(\lambda)\rangle := |U(\lambda)\underline{u}_s(\lambda)\rangle\]are bounded by the product of the magnitudes of the scalar products with an intermediate state.\begin{equation}\label{near-cop} |\langle \underline{\psi}(\lambda_1+\lambda_2) |\underline{\psi}(0)\rangle|\leq |\langle \underline{\psi}(\lambda_1+\lambda_2) |\underline{\psi}(\lambda_1)\rangle |\;| \langle \underline{\psi}(\lambda_1) |\underline{\psi}(0)\rangle|\end{equation}for normalized (\ref{n-minpak-normd}), nonrelativistic classical particle states $|\underline{u}_s(\lambda)\rangle$ from (\ref{n-minpak}), and $\lambda_1,\lambda_2\geq 0$.\end{quote}

Satisfaction of this bound is conditional for the nonrelativistic approximations to the scalar products (\ref{scalarprod}) in this single Lorentz scalar field UQFT example but the bound applies in cases of interest. Conditional satisfaction of the bound is demonstrated for two body circular orbit cases in Section \ref{sec-twop}.

The coplanar propagation bound (\ref{near-cop}) and the unitary implementation of time translation provides that\[ |\langle \underline{u}_s(\lambda_1\!+\!\lambda_2) |U(-(\lambda_1\!+\!\lambda_2))\underline{u}_s(0)\rangle|\leq |\langle \underline{u}_s(\lambda_1\!+\!\lambda_2) |U(-\lambda_2)\underline{u}_s(\lambda_1)\rangle |\;| \langle \underline{u}_s(\lambda_1) |U(-\lambda_1)\underline{u}_s(0)\rangle|.\]As a consequence, when valid, the coplanar propagation bound provides that (\ref{idea}) is satisfied for intervals $(0,\lambda_1+\lambda_2)$ only if it is satisfied for intervals $(0,\lambda_1)$ and $(\lambda_1,\lambda_1+\lambda_2)$. The association of $1/r$ potentials with UQFT over short intervals from domain \ref{item-1} provides that the likelihoods for the trajectories of $1/r$ potentials have the greatest upper bounds in domains \ref{item-2} and \ref{item-3}. That is, the likelihood of a transition along a trajectory cannot be large unless the likelihood is large for intermediate transitions along the same trajectory. The long interval transition likelihoods are bounded by products of the short interval transition likelihoods. Since the trajectories of $1/r$ potentials are the most likely trajectories for a short interval, only the trajectories of a $1/r$ potential can have near certain likelihoods over long intervals. Conversely, validity of this bound provides that the intermediate transition likelihoods along a trajectory must exceed the long interval likelihoods. The likelihood of a transition for a short interval along a trajectory is large if the likelihood of a transition across a larger interval on the same trajectory is large. As a consequence, the potential strength $g$ from domain \ref{item-2} that maximizes the longer interval likelihoods provides that the short interval likelihoods with that strength $g$ must be among the trajectories of great likelihood. Validity of the coplanar propagation bound provides the association of a $-g/r$ potential with the UQFT. A $1/r$ potential is indicated from the short interval results of domain \ref{item-1}, and the strength $g$ is determined from maximization of likelihoods in domain \ref{item-2}. In domain \ref{item-3}, the $-g/r$ potential is indicated as the most likely from among the trajectories of potentials that provide local extrema of the likelihoods.

There are no particle trajectories in quantum mechanics other than the result of a sequence of localized observations that can be reliably associated with a single particle. The coplanar propagation bound relates the likelihood of a trajectory for a particle under recurring observation with the likelihood of observation of the particle at the location and momentum of the result of the series of observed transitions. Each observation has inherent uncertainty and as a consequence, with varying likelihoods, observations associate with many classical trajectories. 

The span of three vectors $|\underline{\psi}(0)\rangle$, $|\underline{\psi}(\lambda_1)\rangle$, $|\underline{\psi}(\lambda_1+\lambda_2))\rangle$ is an at most three dimensional Hilbert subspace [\ref{reed}]. Angles can be associated with the magnitudes of the scalar products.\[\renewcommand{\arraystretch}{1.25}\begin{array}{rl} \cos \theta_1 &=| \langle \underline{\psi}(\lambda_1) |\underline{\psi}(0)\rangle|\\
 \cos \theta_2 &=|\langle \underline{\psi}(\lambda_1+\lambda_2) |\underline{\psi}(\lambda_1)\rangle |\\
 \cos \theta &=|\langle \underline{\psi}(\lambda_1+\lambda_2) |\underline{\psi}(0)\rangle|\end{array}\]with $0\leq \theta_1,\theta_2,\theta \leq \pi/2$. In those cases with $\theta_1=\theta_2$, the coplanar propagation bound is that\[\cos^2\theta_1\geq \cos \theta.\]When $\theta_1=\theta_2$, the angle $\beta$ of the intermediate state $|\underline{\psi}(\lambda_1)\rangle$ out of the plane determined by the initial and final states is, from the law of cosines,\[\cos^2 \beta = \frac{2 \cos^2\theta_1}{1+\cos\theta}\geq \frac{2 \cos\theta}{1+\cos\theta}.\]The coplanar propagation bound is a bound on the deviation of the intermediate state from the plane of the initial and final states. Asymptotically for small $\theta$, the coplanar propagation bound is that $\beta \leq \theta /2$ when $\theta_1=\theta_2$. If the three vectors are coplanar, then $\beta=0$ and $\theta_1=\theta/2$.

The results introduced as \ref{item-1}-\ref{item-4} are developed in the subsequent sections after the UQFT VEV are introduced.

\subsection{The single Lorentz scalar field Wightman functions}\label{sec-vev}

A useful characterization of the Wightman functions $W_n((x)_n)$ decomposes the generalized functions into connected functions ${^CW}_n((x)_n)$. Connected functions exhibit the {\em cluster} property and decline as the support of arguments separate. ${^CW}_{2n}(s_n^* \,s_n)$ is negligible if for a $j\neq k$, $\|\xi_j-\xi_k\|$ increases without bound with $s_n$ a product (\ref{n-minpak}) of Gaussian minimum packet functions centered on locations $\xi_k$. Within the original Wightman axioms, this decomposition is the link-cluster expansion [\ref{bogo},\ref{ruelle}] and the connected functions ${^CW}_n((x)_n)$ are known as truncated VEV. The link-cluster expansion results when there is a Hermitian field and a distinct link-cluster-like expansion applies for UQFT [\ref{intro}]. In UQFT with a single Lorentz scalar field, the two-point connected VEV are Pauli-Jordan positive frequency (generalized) functions.\[{^C\tilde{W}}_2(p_1,p_2):= \delta({\bf p}_1+{\bf p}_2)\, 2\omega_1 \delta(p_1^2-m^2)\theta(-E_1)\delta(p_2^2-m^2)\theta(E_2),\]a convenient equivalent to the more familiar representation of the Pauli-Jordan function as the Fourier transform of\[{^C\tilde{W}}_2(p_1,p_2)= \theta(E_2) \delta(p_2^2-m^2) \delta(p_1+p_2).\]

The VEV for the selected UQFT are inverse Fourier transforms of elementary generalized functions [\ref{gel2}]. The forward contribution to the $2n$ point VEV consists of a product of two-point connected functions with momenta equal in pairs. Limited to ${\cal B}$, the forward contribution to the $2n$ point VEV coincides with the $\tilde{W}_{2n}((p)_{2n})$ of a free field.\begin{equation}\label{vev-forw} {^F}\tilde{W}_{2n}((p)_{2n})=\sum_{\mathit{pairs}} \prod_{k=1}^n 2\omega_k \delta({\bf p}_k-{\bf p}_{i_k})\,\delta(p_k^2-m^2)\theta(-E_k)\delta(p_{i_k}^2-m^2)\theta(E_{i_k})\end{equation}with the summation over the $n!$ distinct pairs $(1,i_1),\ldots (n,i_n)$ with $i_k\in \{n+1,\ldots 2n\}$. The connected contributions to the $n$ point VEV $W_n((x)_n)$ for $2n\geq 4$ determine the interaction and have the Fourier transforms\begin{equation}\label{vev-connctd} {^C\tilde{W}}_n((p)_n)=c_n\, \delta(p_1+p_2\ldots + p_n)\prod_{k=1}^n \delta(p_k^2-m^2).\end{equation}At least three spacetime dimensions are required for (\ref{vev-connctd}) to define a generalized function [\ref{intro},\ref{gej05}], and four dimensions are required to include $m=0$ [\ref{mp01}]. More generally for a single Lorentz scalar field, ${^C\tilde{W}}_{2n}((p)_{2n})$ includes a factor $Q_{n,2n}((p)_{2n})$ of a function symmetric under interchange of any two momentum arguments $p_k$ and $p_j$ when either $1\leq k,j \leq n$ or $n+1\leq k,j \leq 2n$. Examples of $Q_{n,2n}((p)_{2n})$ are in [\ref{gej05},\ref{feymns}]. $c_n$ sets the relative strength of the connected contributions with respect to the forward contributions. The $c_n$ are the moments of a nonnegative measure, and $c_n$ has the units of (length)$^{n-4}$. For four and six arguments,\[\renewcommand{\arraystretch}{1.75} \begin{array}{rl} W_4((x)_4)&={^CW}_4((x)_4)+{^CW}_2(x_1,x_3){^CW}_2(x_2,x_4)+{^CW}_2(x_1,x_4){^CW}_2(x_2,x_3)\\
W_6((x)_6) &={^CW}_6((x)_6)+\frac{1}{9}\;{\ds \sum_{\mathit{pairs}} \sum_{k=1}^3}\; {^CW}_2(x_k,x_{i_k}) {^CW}_4(x_{[k+1]},x_{[k+2]},x_{i_{[k+1]}},x_{i_{[k+2]}})\\
 &\qquad\qquad +{\ds \sum_{\mathit{pairs}}}\; {^CW}_2(x_1,x_{i_1}){^CW}_2(x_2,x_{i_2}){^CW}_2(x_3,x_{i_4}).\end{array}\]These expressions apply in ${\cal B}$ and expressions that encompass ${\cal A}$ are in [\ref{intro}]. The combinatoric factor $1/9$ and the lack of contributions from terms with factors such as ${^CW}_2(x_1,x_2)$ distinguish the UQFT expansion from the link-cluster expansion in Wightman QFT. Here, $[j]\in\{1,2,3\}$ and is equivalent to $j$ modulo 3. The summation over pairs consists of all selections for $i_1,i_2,i_3$ that are permutations of $\{4,5,6\}$ without regard to order: the 3 distinct pairings $(k,i_k)$ for the product of a two-point and a four-point connected function, and $3!$ distinct pairs $(1,i_1)(2,i_2)(3,i_3)$ for the product of three two-point functions.

An abbreviated notation introduced in Section \ref{sec-cllim} is used for the decomposition of scalar products into forward, partially factored, and connected contributions.\begin{equation}\label{expand} F(\lambda,\tau)+c_{2n}C(\lambda,\tau):=\langle U(\lambda)\underline{s}(\lambda)|U(\tau)\underline{s}(\tau)\rangle\end{equation}with the designations $F(\lambda,\tau)$ for all forward VEV contributions and $c_{2n}C(\lambda,\tau)$ for all connected VEV contributions. $F+c_{2n}C$ is an abbreviated notation to describe the full link-cluster-like expansion and the partially factored contributions are implicit. From (\ref{vev-forw}), the forward contributions to the scalar product are\begin{equation}\label{S-f-defn} F(\lambda,\tau):= {\ds \int} d(x)_{2n}\; {^FW}_{2n}((x)_{2n}) s_n((x-\lambda)_n;\lambda)^*s_n((x-\tau)_{n+1,2n};\tau)\end{equation}with $s_n((x)_n;\tau)\in {\cal B}$ and\[(x-\lambda)_{j,k}=t_j-\lambda,{\bf x}_j,\ldots, t_k-\lambda,{\bf x}_k.\]From (\ref{vev-connctd}), the connected contributions are\begin{equation}\label{S-c-defn} c_{2n}C(\lambda,\tau):= {\ds \int} d(x)_{2n}\; {^CW}_{2n}((x)_{2n}) s_n((x-\lambda)_n;\lambda)^*s_n((x-\tau)_{n+1,2n};\tau).\end{equation}

\section{States with classical descriptions}

Gaussian minimum packet functions and the trajectories of nonrelativistic classical particles are described in this section. Elements of ${\cal B}$ correspond to products of Gaussian minimum packet functions and label states of a UQFT. Trajectories that are solutions to Newton's equations for $1/r$ potentials naturally associate with Gaussian minimum packet functions.

\subsection{Classical, nonrelativistic trajectories}\label{nonrel}

Classical trajectories $\xi_k(\lambda)$ are functions from a real temporal parameter $\lambda$ to an ${\bf R}^3$ configuration space. The $\xi_k(\lambda)$ have units of length and $k$ labels one of a set of $n$ trajectories. To establish a nonrelativistic classical correspondence, classical trajectories with nonrelativistic relative velocities are necessary to diminish the likelihood of particle creation or annihilation and achieve a persistent correspondence of relativistic quantum physics with classical particles. To identify dominant regions in the support of the functions describing the quantum states as classical objects requires both that particle number changes are negligible and that the dominant regions remain isolated.

Momenta ${\bf q}_k$ are related to the trajectories by\begin{equation}\label{q-defn}{\bf q}_k(\lambda):= \frac{1}{\lambda_c}\,\gamma_k\, \dot{\xi}_k(\lambda)\end{equation}with $\lambda_c$ the Compton wavelength (\ref{compton}) and\[\gamma_k :=\frac{1}{\sqrt{1-\dot{\xi}_k^2}}.\]$\dot{\bf \xi}_k(\lambda)$ is the first derivative with respect to $\lambda$ of ${\bf \xi}_k(\lambda)$. Energies are $\omega({\bf q}_k)$ from (\ref{omega}) and then $q_k^2=\lambda_c^{-2}$.

In the nonrelativistic limit, relative velocities are much less than the speed of light and in coordinates collocated with the center of mass, $\dot{\xi}_k(\lambda)^2\ll 1$. As a consequence, to first contributing order in the $\dot{\xi}_k$,\[{\bf q}_k \approx \frac{1}{\lambda_c} \dot{\xi}_k\]and\begin{equation}\label{nr-approx}\omega({\bf q}_k)\approx \frac{1}{\lambda_c}(1+\frac{\dot{\xi}_k^2}{2}).\end{equation}When $\dot{\xi}_k^2\ll 1$, $(\hbar {\bf q}_k)^2\ll (mc)^2$. 

The classical potential energy of an assembly of $n$ particles is\begin{equation}\label{potnl}mc^2V(\lambda):=\sum_{j=1}^n \sum_{k >j}^n \Phi_{jk}(\|\xi_j(\lambda)-\xi_k(\lambda)\|)\end{equation}with $\Phi_{jk}(r)=\Phi_{k j}(r)$ the pair potential. Below, it is convenient to express results in terms of this potential energy and\begin{equation}\label{it-defn}\renewcommand{\arraystretch}{1.25}\begin{array}{rl} I(\lambda) &:= {\ds \frac{1}{2}}\; {\ds \sum_{k=1}^n} \xi_k(\lambda)^2\\
 T(\lambda) &:= {\ds \frac{1}{2}}\; {\ds \sum_{k=1}^n} \dot{\xi}_k(\lambda)^2.\end{array}\end{equation}$mI$ is the scalar moment of inertia about the origin and $mc^2 T$ is the nonrelativistic kinetic energy of the assembly of $n$ classical particles all of mass $m$. The classical quantities $I$, $T$ and $V$ are functions of the trajectories $\xi_k(\lambda)$ or first derivatives $\dot{\xi}_k(\lambda)$. 

Newton's equation of motion is\begin{equation}\label{newton} \ddot{\xi}_k= -\frac{\partial V}{\partial \xi_k}.\end{equation}For trajectories $\xi_k(\lambda)$ that satisfy Newton's equation, the classical total energy\begin{equation}\label{totE}e_C:=\frac{E}{mc^2} =T(\lambda)+V(\lambda)\end{equation}is a constant of the motion.

Total momentum is a constant of the motion for the set of trajectories $\xi_k(\lambda)$ [\ref{goldstein}]. From the Poincar\'{e} covariance of the UQFT, there is no loss of generality from selection of an inertial coordinate frame collocated with the classical center of mass. In this frame,\[\sum_{j=1}^n \xi_j(\lambda) = \sum_{j=1}^n \dot{\xi}_j(\lambda)=0\]for all $\lambda$.

\subsection{Minimum packet states}\label{sec-minpak}


In ordinary, nonrelativistic quantum mechanics, states are described by equivalence classes of square-summable functions, time is an independent parameter, and position ${\bf X}$ and momentum ${\bf P}$ are Hermitian Hilbert space operators realized in ${\cal L}_2({\bf R}^3)$ by ${\bf X}={\bf x}$ and ${\bf P}=-i\hbar \nabla$. In ordinary quantum mechanics, minimum packet functions label quantum states with descriptions that are most nearly classical in the sense that the states have a simultaneously determined position and momentum to the greatest degree consistent with nature. For minimum packet functions, the geometric means of the variances in positions and momenta achieve the Heisenberg bound.

A classical trajectory $\xi_k(\lambda)$ describes a family of minimum packet functions; a family parameterized by the temporal parameter $\lambda$. These functions are peaked on positions and momenta described by the classical trajectory $\xi_k(\lambda)$. The Fourier transforms of this family of minimum packet functions are Schwartz tempered functions\begin{equation}\label{minpak-p} \tilde{\varphi}({\bf p};{\bf \xi}_k,\lambda):=\frac{L_0^3}{\sqrt{\pi^3}}\,\exp\left(-\ell_0^2 ({\bf p}-{\bf q}_k)^2+i ({\bf p}-{\bf q}_k)\!\cdot\!\xi_k\right).\end{equation}The complex parameter\[\ell_0^2:=L_0(\lambda)^2-i\frac{\lambda_c}{2} \lambda\]is a function of $\lambda$. $L_0$ is abbreviated notation for the real, positive definite function $L_0(\lambda)$, and $\lambda_c$ is the Compton wavelength (\ref{compton}). ${\bf q}_k$ is the momentum from (\ref{q-defn}). States are rays in the Hilbert space and (\ref{minpak-p}) is a convenient selection for magnitude and phase. The Gaussian minimum packet functions are the three dimensional inverse Fourier transforms of (\ref{minpak-p}),\begin{equation} \label{minpak} \varphi({\bf x}):=\varphi({\bf x};{\bf \xi}_k,\lambda)=\frac{1}{\sqrt{(2\pi)^3}}\frac{L_0^3}{\ell_0^3}\; \exp\left(-\frac{({\bf x}-{\bf \xi}_k)^2}{4\ell_0^2}-i {\bf q}_k\!\cdot\!{\bf x} \right).\end{equation}With the designation $\langle T \rangle_2 := N^{-1} \int d{\bf x}\; \overline{\varphi({\bf x})}\, T\varphi({\bf x})$ for an ${\cal L}_2$ operator $T$, the normalization $N:=\int d{\bf x}\; |\varphi({\bf x})|^2$ results in\[\renewcommand{\arraystretch}{1.25} \begin{array}{l} \langle X\rangle_2= \xi_k\\
\langle P\rangle_2= {\bf q}_k\\
\sigma_X^2=\langle (X-\langle X\rangle_2 )^2 \rangle_2= L_0^2+\lambda_c^2 \lambda^2/4L_0^2\\
\sigma_P^2=\langle (P-\langle P\rangle_2)^2 \rangle_2= 1/4L_0^2\\
\sigma_X \sigma_P ={\ds \frac{1}{2}}\sqrt{1+\lambda_c^2 \lambda^2/4L_0^4}, \end{array}\]the minimum consistent with the Heisenberg uncertainty at $\lambda=0$. $L_0$ increasing without bound provides a delta sequence.\begin{equation}\label{delta-seq}\delta_{L_0}(s):= \frac{L_0}{\sqrt{\pi}}\,e^{-L_0^2s^2}\approx \delta(s).\end{equation}These delta sequences approximate {\em plane wave states} that are generalized eigenvectors of the UQFT Hamiltonian (\ref{hamil}) but not elements of the Hilbert space.

Elements of the Hilbert space are labeled by functions from ${\cal B}$. The products of minimum packet functions (\ref{minpak-p}) and $E_k+\omega_k$ are the functions from ${\cal B}$ used to describe the classical, particle states in (\ref{n-minpak}). From (\ref{B-defn}) and $\tilde{\varphi}_n(({\bf p})_n) \in {\cal S}({\bf R}^{3n})\subset {\cal A}$, Schwartz test functions of momenta generate functions in ${\cal B}$. $E_k$ and $\omega_k$ are multipliers [\ref{gel2}] of spacetime Schwartz functions. The spatial support of Gaussian minimum packet functions is concentrated in a neighborhood of $\xi_k(\lambda)$ of extent proportional to the square root of $L_0^2+\lambda_c^2 \lambda^2/4L_0^2$. This support can be arbitrarily dominantly within a small neighborhood when $L_0$ and $\lambda$ are small, a result that persists with multiplication by $E_k+\omega_k$. Multiplication by $E_k$ corresponds to a differentiation with respect to time. Multiplication by $\omega_k$ transforms a spacetime Schwartz function of bounded spatial support to an anti-local function, a function that does not vanish in any spatial neighborhood [\ref{reeh},\ref{segal},\ref{masuda}]. Nevertheless, a minimum packet function dominantly supported in a small neighborhood remains dominantly supported in a small neighborhood after multiplication by $\omega_k$ due to the rapid decline of Gaussian functions.

In [\ref{limits}], it is demonstrated that to first contributing order in small $\lambda$ and for nonrelativistic relative velocities, temporal translations of the UQFT states are equivalent to translations of the support of minimum packet function labels. These translations are along classical trajectories that satisfy Newton's equation (\ref{newton}) for a $1/r$ potential. A stronger statement of the result is:
\newline

\newcounter{theorems}
\setcounter{theorems}{1}
\renewcommand{\thetheorems}{\arabic{theorems}}
{\bf Lemma} \thetheorems: For the $n$ particle minimum packet functions (\ref{minpak-p}) with momentum spread lengths $L_0(\lambda)$ that are continuous functions of $\lambda$ and in the nonrelativistic limit, the dominant regions in the support of the functions traverse the classical trajectories of a $1/r$ potential asymptotically at small $\lambda$,\[ |\underline{s}(0)\rangle = | U(\lambda) \underline{s}(\lambda)\rangle.\]
\newline
The demonstration follows from approximation of the functions labeling the state $| U(\lambda) \underline{s}(\lambda)\rangle$. If\[ e^{-i\omega_k \lambda} \tilde{\varphi}({\bf p};{\bf \xi}_k,\lambda)e^{i\phi_k(\lambda)} \approx \tilde{\varphi}({\bf p};{\bf \xi}_k,0)\]to first order in $\lambda$, then the lemma results from the definition (\ref{n-minpak}) of the functions that label the states $|\underline{s}(\lambda)\rangle$, the form of the UQFT Hamiltonian (\ref{hamil}), and that equality as states allows a phase difference. Due to the limitation of the support of the VEV to mass shells and the rapid decline of the minimum packet functions, multiplication by $E_k+\omega_k$ is multiplication by $2\omega_k\approx 2/\lambda_c$ in the nonrelativistic limit and the indicated equality of minimum packet functions suffices to demonstrate the lemma.

The inverse Fourier transform of the time translated minimum packet function (\ref{minpak-p}) multiplied by the unimodular factor from a selected phase $\phi_k(\lambda)$ is\begin{equation}\label{lem1-eq1}\renewcommand{\arraystretch}{2.25}\begin{array}{l} U(\lambda)\varphi({\bf x};{\bf \xi}_k,\lambda)e^{i\phi_k(\lambda)} = {\ds \int}{\ds \frac{d{\bf p}}{(2\pi)^\frac{3}{2}}}\; e^{-i{\bf p}\cdot{\bf x}}e^{-i\omega({\bf p}) \lambda}\tilde{\varphi}({\bf p};{\bf \xi}_k,\lambda)e^{i\phi_k(\lambda)}\\
 \qquad\qquad = {\ds \frac{L_0^3}{(2\pi^2)^\frac{3}{2}}}{\ds \int}d{\bf p}\; e^{-i{\bf p}\cdot{\bf x}-i\omega({\bf p}) \lambda -\ell_0^2 ({\bf p}-{\bf q}_k)^2+i ({\bf p}-{\bf q}_k)\cdot\xi_k+i\phi_k(\lambda)}\\
 \qquad\qquad \approx {\ds \frac{L_0^3}{(2\pi^2)^\frac{3}{2}}}e^{-i{\bf q}_k\cdot{\bf x}} {\ds \int}d{\bf p}'\; e^{-i{\bf p}'\cdot{\bf x} -i\frac{1}{2 \lambda_c}\dot{\xi}_k^2\lambda -i{\bf p}'\cdot \dot{\xi}_k\,\lambda -L_0^2 {\bf p}'^2+i {\bf p}'\cdot\xi_k + i\frac{\lambda}{n\lambda_c}e_C}\\
 \qquad\qquad = {\ds \frac{1}{(2\pi)^\frac{3}{2}}} \exp\left( -{\ds \frac{({\bf x}-\xi_k+\lambda \dot{\xi}_k)^2}{4 L_0^2}}\right) e^{-i\frac{1}{\lambda_c} \dot{\xi}_k\cdot{\bf x}-i\frac{1}{2 \lambda_c}\dot{\xi}_k^2\lambda + i\frac{\lambda}{n\lambda_c}e_C}\end{array}\end{equation}with $\xi_k$ and $L_0$ evaluated at $\lambda$. This result used a change of summation variable to ${\bf p}':={\bf p}-{\bf q}_k$, a substitution for $\omega_k$ valid for nonrelativistic momenta, a substitution for $\phi_k(\lambda)$, the relation (\ref{q-defn}) between ${\bf q}_k$ and $\dot{\xi}_k$, and the Gaussian summation\begin{equation}\label{gauss-sum}\int_{-\infty}^{\infty} ds\; e^{-\alpha s^2 +\beta s} = \sqrt{\pi}\;\frac{e^{\beta^2/(4 \alpha)}}{\sqrt{\alpha}}\end{equation}for $\Re e(\alpha)>0$. When the classical trajectory is nonrelativistic and the momentum support of the packet is sufficiently limited near ${\bf q}_k$ that\begin{equation}\label{nr-p-approx}\omega_k \approx \omega({\bf q}_k) +\frac{{\bf q}_k\cdot ({\bf p}_k-{\bf q}_k)}{\omega({\bf q}_k)}+\frac{({\bf p}_k-{\bf q}_k)^2}{2\omega({\bf q}_k)},\end{equation}then the nonrelativistic approximation\[\omega({\bf q}_k) \approx {\ds \frac{1}{\lambda_c}} + {\ds \frac{\dot{\xi}_k^2}{2\lambda_c}},\]from (\ref{nr-approx}) and the selection of phase\[\phi_k(\lambda):= \frac{1}{\lambda_c}\,(1 + \frac{e_C}{n})\, \lambda\]result in the expression (\ref{lem1-eq1}) above. The classical energy $e_C$ from (\ref{totE}) is a constant for trajectories that satisfy Newton's equation (\ref{newton}). Justification of the nonrelativistic expansion (\ref{nr-p-approx}) and a demonstration that (\ref{lem1-eq1}) is equal to $\varphi({\bf x};{\bf \xi}_k,0)$ to first order in $\lambda$ complete the lemma.

(\ref{nr-p-approx}) derives from Taylor's theorem that results in the approximation\[\omega_k \approx \omega({\bf q}_k) + \frac{{\bf q}_k\cdot ({\bf p}_k-{\bf q}_k)}{\omega({\bf q}_k)} +\frac{({\bf p}_k-{\bf q}_k)^2}{2\omega({\bf q}_k)} -\frac{({\bf q}_k\cdot({\bf p}_k-{\bf q}_k))^2}{2\omega({\bf q}_k)^3}\]and the Cauchy-Schwarz-Bunyakovski inequality provides that\[({\bf q}_k\cdot({\bf p}_k-{\bf q}_k))^2\leq {\bf q}_k^2\, ({\bf p}_k-{\bf q}_k)^2.\]For nonrelativistic classical velocities\[{\bf q}_k^2 \ll \omega({\bf q}_k)^2\]and the second term in the second order correction in the Taylor series is negligible for the nonrelativistic approximation. From\[{\bf p}_k^2=({\bf p}_k-{\bf q}_k +{\bf q}_k)^2=({\bf p}_k-{\bf q}_k)^2- 2{\bf q}_k\cdot ({\bf p}_k-{\bf q}_k)+{\bf q}_k^2,\]a bound on the error in (\ref{nr-p-approx}) is\begin{equation}\label{taylor} \left|\, \omega_k -\omega({\bf q}_k) -\frac{{\bf q}_k\cdot ({\bf p}_k-{\bf q}_k)}{\omega({\bf q}_k)}-\frac{({\bf p}_k-{\bf q}_k)^2}{2\omega({\bf q}_k)}\right| \leq \frac{\lambda_c^3}{4} ({\bf p}_k^2-{\bf q}_k^2)^2.\end{equation}Then, ${\bf p}_k^2-{\bf q}_k^2=({\bf p}_k-{\bf q}_k)\cdot({\bf p}_k+{\bf q}_k)$, $1\leq \lambda_c \omega({\bf q}_k)$, the Cauchy-Schwarz-Bunyakovski inequality and the parallelogram law result in the convenient bound\[\frac{1}{4} ({\bf p}_k^2-{\bf q}_k^2)^2 \leq \frac{1}{2} ({\bf p}_k-{\bf q}_k)^4 +2{\bf q}_k^2 \, ({\bf p}_k-{\bf q}_k)^2 \]in terms of $\|{\bf p}_k-{\bf q}_k\|$. The error vanishes at the peak of the momentum support of the minimum packet. When\[\|{\bf p}_k-{\bf q}_k\| \ll \frac{1}{\lambda_c}\leq \omega_k,\]the error in (\ref{nr-p-approx}) is negligible with respect to $\omega_k$. The dominant support of the minimum packet is in a neighborhood of ${\bf q}_j$ of scale proportional to $1/L_0$ and the nonrelativistic approximation is accurate when $L_0$ is bounded from below.\[ \|{\bf p}_k-{\bf q}_k\| \sim \frac{1}{L_0} \ll \frac{1}{\lambda_c}\]or\begin{equation}\label{pckt-size}\lambda_c \ll L_0.\end{equation}

Equality of (\ref{lem1-eq1}) with $\varphi({\bf x};{\bf \xi}_k,0)$ follows for the particular choice of a $1/r$ potential,\begin{equation}\label{r-potnl}\Phi_{kj}(\|\xi_k-\xi_j\|)= \frac{-g}{\|\xi_k-\xi_j\|}.\end{equation}A property of the $1/r$ potential, equal and opposite actions, the definition (\ref{potnl}) of $V$ and Newton's equation (\ref{newton}) provides that\[V={\ds \sum_{k=1}^n} \ddot{\xi}_k \!\cdot\! \xi_k.\]This relation follows from the chain rule for derivatives applied to the potential (\ref{potnl}) with the substitution of the $1/r$ potential (\ref{r-potnl}). From\[ {\ds \frac{g}{r}} =-\left(\nabla {\ds \frac{g}{r}}\right)\cdot \xi =-\left({\ds \frac{\partial \;}{\partial r}} {\ds \frac{g}{r}}\right) \left( {\ds \frac{\partial r}{\partial \xi_x}}\xi_x+{\ds \frac{\partial r}{\partial \xi_y}}\xi_y+{\ds \frac{\partial r}{\partial \xi_z}}\xi_z\right)\]with $r:=\|\xi\|$, it follows that\[\renewcommand{\arraystretch}{2.25}\begin{array}{rl} V &=-{\ds \sum_{k=1}^n \sum_{j >k}^n} {\ds \frac{\partial \Phi_{kj}(\|\xi_k-\xi_j\|)}{\partial(\xi_k-\xi_j)}}\cdot (\xi_k-\xi_j)\\
 &=-{\ds \sum_{k=1}^n \sum_{j >k}^n} \left( {\ds \frac{\partial \Phi_{kj}(\|\xi_k-\xi_j\|)}{\partial \xi_k}}\cdot \xi_k +{\ds \frac{\partial \Phi_{kj}(\|\xi_k-\xi_j\|)}{\partial \xi_j}}\cdot \xi_j \right)\\
 &=-{\ds \sum_{k=1}^n \sum_{j\neq k}^n} {\ds \frac{\partial \Phi_{kj}(\|\xi_k-\xi_j\|)}{\partial \xi_k}}\cdot \xi_k\\
 &=-{\ds \sum_{k=1}^n} {\ds \frac{dV}{d\xi_k}}\!\cdot \! \xi_k.\end{array}\]Substitution of Newton's equation of motion (\ref{newton}) results in the desired identity. An upper bound on $L_0$ permits the approximation\begin{equation}\label{upr-bnd} \renewcommand{\arraystretch}{1.25}\begin{array}{rl} V&={\ds \sum_{k=1}^n} \ddot{\xi}_k \!\cdot\! {\bf x}+{\ds \sum_{k=1}^n} \ddot{\xi}_k \!\cdot\! (\xi_k-{\bf x})\\
 &\approx {\ds \sum_{k=1}^n} \ddot{\xi}_k \!\cdot\! {\bf x}\end{array}\end{equation}with no error at the peak of the support the minimum packet function.

The product over the approximations (\ref{lem1-eq1}) to $n$ minimum packet functions includes unimodular factors from the selection for phases $\phi_k(\lambda)$ . These factors result in\[\renewcommand{\arraystretch}{2.25}\begin{array}{rl} {\ds \prod_{k=1}^n} e^{i\frac{\lambda}{\lambda_c n} e_C} &= e^{i\frac{\lambda}{\lambda_c}e_C}\\ 
 &= {\ds \prod_{k=1}^n} e^{i \frac{\lambda}{2\lambda_c} \dot{\xi}_k(\lambda)^2 + i\frac{\lambda}{\lambda_c} \ddot{\xi}_k \cdot {\bf x}}\end{array}\]from the substitutions\[e_C=\sum_{k=1}^n \frac{\dot{\xi}_k^2}{2} +V\]from the definition of kinetic energy (\ref{it-defn}) and total energy (\ref{totE}), and $V$ from (\ref{upr-bnd}). The indicated substitutions in the final line of (\ref{lem1-eq1}) in the corresponding factor for each minimum packet function labeled $k$ in (\ref{n-minpak}) results in\[U(\lambda)\varphi({\bf x};{\bf \xi}_k,\lambda)e^{i\phi_k(\lambda)} \approx {\ds \frac{1}{\sqrt{(2\pi)^3}}}\,\exp\left( -{\ds \frac{({\bf x}-\xi_k+\lambda \dot{\xi}_k)^2}{4 L_0^2}}\right) e^{-i\frac{1}{\lambda_c} (\dot{\xi}_k-\ddot{\xi}_k \lambda) \cdot {\bf x}}\]with $\xi_k$, $\dot{\xi}_k$, $\ddot{\xi}_k$ and $L_0$ evaluated at $\lambda$. For sufficiently small $\lambda$,\[\xi_k(0)\approx \xi_k(\lambda)-\lambda \dot{\xi}_k(\lambda)\]from Taylor's theorem. This substitution results in the asymptotic equality at small $\lambda$ of the inverse Fourier transforms of\[e^{-i\omega_k \lambda}\tilde{\varphi}({\bf p}_k;\xi_k,\lambda)e^{i\phi_k(\lambda)} \approx \tilde{\varphi}({\bf p}_k;\xi_k,0)\]in the nonrelativistic, classical particle approximation and when $L_0(\lambda)$ is constant. Variation of $L_0$ with $\lambda$ modifies the extent of the packet but not the location of the peak. As a consequence, a continuous $L_0(\lambda)$ suffices for the result to apply asymptotically for small $\lambda$ to the trajectory of the dominant support of the minimum packet function and demonstrates the lemma. 

To second order in the nonrelativistic approximation to the energies $\omega_k$, the imaginary contribution to $\ell_0$ in $\tilde{\varphi}({\bf p}_k;{\bf \xi}_k,\lambda)$ is removed by the time translation $U(\lambda)$ and a Gaussian packet shape (\ref{minpak-p}) is preserved by time translation.

These considerations put both upper and lower bounds on $L_0$. The lower bound (\ref{pckt-size}) implements a nonrelativistic approximation and the accuracy of the approximation (\ref{upr-bnd}) is satisfied if there is an appropriate upper bound on $L_0$. Newton's equation (\ref{newton}), the $1/r$ potential (\ref{r-potnl}), and the Cauchy-Schwarz-Bunyakovski inequality provides that\[|\ddot{\xi}_k\cdot ({\bf x}-\xi_k)|\leq \|\ddot{\xi}\|\, \|{\bf x}-\xi_k \|<\frac{g}{r_a^2} \,k_0\, \sigma_X\]with $r_a$ the closest approach of any two trajectories at the time $\lambda$ of interest. The support of the minimum packet functions is dominantly within a neighborhood $\|{\bf x}-\xi_k\|<k_0 \,\sigma_X$ with $3<k_0<10$ typically sufficing. An estimate for the error in the approximations (\ref{upr-bnd}) results from the dominant term of the potential energy $V$, the term for the pair at the closest approach.\[|\ddot{\xi}_k\cdot ({\bf x}-\xi_k)|\leq \frac{g}{r_a^2} k_0\, \sigma_X \ll \frac{g}{r_a}.\]This is equivalent to\[\sqrt{L_0^2+\lambda_c^2 \lambda^2/4L_0^2}=\sigma_X \ll r_a\]and provides that to high likelihood, the trajectories are identifiable and remain distinct. The lower bound on $L_0$ (\ref{pckt-size}) is combined with the upper bound from satisfaction of (\ref{upr-bnd}) to determine the minimum packet functions that correspond to nonrelativistic classical particles. The result is one of the nonrelativistic classical particle conditions (\ref{L0-ineq}) of Section \ref{sec-defn}.

The equivalence of small temporal translations with translation of corresponding particles along the classical trajectories is independent of the VEV. The $1/r$ potential is associated with the minimum packet functions independently of the VEV. This small $\lambda$ result is a property of minimum packet states and that the quantum states can be localized in time. This localization applies to UQFT that include states labeled by minimum packet functions within ${\cal B}$ with point support in time.

\section{Scalar products of minimum packet UQFT states}\label{sec-eval}

In this section, the scalar products (\ref{isometry}) of states described by time translated, $n$ particle, minimum packet states (\ref{n-minpak}) are evaluated. The scalar products of interest (\ref{scalarprod}) in evaluation of Born's rule transition amplitudes (\ref{likelihood}) derive from\[\langle U(\lambda) \underline{s}(\lambda)|\underline{s}(0) \rangle \]with $\lambda \geq 0$. The growth in the uncertainty in location implied by an initial uncertainty in momentum generally increases with time and establishes a direction to time. The scalar product is evaluated for states associated with classical trajectories that propagate forward in time from an initial state. 

In the nonrelativistic approximation, the apparently more general $\langle U(\lambda) \underline{s}(\lambda) | U(\tau) \underline{s}(\tau) \rangle$ results from $\langle U(\lambda) \underline{s}(\lambda) | \underline{s}(0) \rangle$ using the unitary implementation of time translation and a translation of the temporal parameter of the classical trajectories. Unitary implementation of time translation provides that\[\langle U(\lambda) \underline{s}(\lambda) | U(\tau) \underline{s}(\tau) \rangle=\langle U(\lambda-\tau) \underline{s}(\lambda) | \underline{s}(\tau) \rangle\]when $\lambda\geq \tau$. Translation invariance of the Wightman functions and a temporal translation in the parameterization of the classical trajectories result in\begin{equation}\label{lmda-trans} \langle U(\lambda-\tau) \underline{s}(\lambda-\tau) | \underline{s}(0)\rangle \mapsto \langle U(\lambda) \underline{s}(\lambda) | U(\tau) \underline{s}(\tau) \rangle\end{equation}using\[\xi_k(\tau')\mapsto \xi_k(\tau'+\tau)\]and that the temporal dependence of the $\underline{s}_n(\lambda)$ outside of the $i\lambda_c \lambda/2$ contribution to $\ell_0^2$ in (\ref{minpak-p}) is due entirely to the $\xi_k(\lambda)$ since from (\ref{L0-form}), the $\lambda$ dependence of $L_0(\lambda)$ derives from the $\xi_k(\lambda)$. The comment following the demonstration of Lemma 1 provides that the imaginary component of $\ell_0^2$ in $\underline{s}_n(\lambda)$ compensates for the time translation $U(\lambda)$, and to comparable accuracy as the nonrelativistic approximation, this dependence can also be translated. The scalar products are determined to satisfy\[\langle U(\lambda) \underline{s}(\lambda) | U(\tau) \underline{s}(\tau) \rangle=\overline{ \langle U(\tau) \underline{s}(\tau) | U(\lambda) \underline{s}(\lambda) \rangle}\]for all $\lambda,\tau$.

From (\ref{expand}) in Section \ref{sec-vev}, the scalar product (\ref{scalarprod}) includes forward and connected contributions $F(\lambda,0)$ and $c_{2n}C(\lambda,0)$. These contributions are evaluated in the following sections.

\subsection{Connected contributions}\label{sec-connct}

The connected contributions $c_{2n}C(\lambda,\tau)$ to the scalar product (\ref{scalarprod}) are evaluated in this section for UQFT VEV and minimum packet functions $\tilde{s}_n((p)_n;\lambda)\in {\cal B}$ centered on classical, $n$ particle trajectories.

The result is developed for a general set of trajectories $\xi_k(\lambda)$ but the statement of the lemma is simplified to apply for trajectories that conserve the classical momentum. The lemma applies in a coordinate frame collocated with the classical center of mass. There is no loss in generality in this selection due to the Poincar\'{e} invariance of the scalar product and classical momentum conservation. Intermediate results include the case of a generally specified set of $({\bf q})_n$ used in the evaluation of plane wave limit scattering amplitudes. 
\newline

\addtocounter{theorems}{1}
{\bf Lemma} \thetheorems: For the connected VEV (\ref{vev-connctd}) and minimum packet functions (\ref{n-minpak}), the connected contributions (\ref{S-c-defn}) to the scalar product (\ref{scalarprod}) are\begin{equation}\label{reslt-connctd}C(\lambda,\tau) = \frac{L_e^3 \, e^{i\theta_C}}{(2\pi n)^2 \sqrt{\widehat{{\bf a}^2}}} \; e^{-n (\widehat{{\bf a}^2}\; \widehat{{\bf b}^2}- \widehat{{\bf a}\cdot{\bf b}}^2)/2 \widehat{{\bf a}^2}} e^{-(\delta T)^2/2n \widehat{{\bf a}^2}}\end{equation}in nonrelativistic classical particle instances (\ref{L0-ineq}) in coordinates collocated with the classical center of mass, and with\begin{equation}\label{defn-connctd}\renewcommand{\arraystretch}{2.25} \begin{array}{rl} \delta T &:={\ds \frac{1}{\lambda_c}} (T(\lambda)-T(\tau))\\
 \theta_C &:=n\, {\ds \frac{1}{\lambda_c}}\, (\lambda-\tau)+ {\ds \frac{1}{\lambda_c}}\, (T(\lambda)\,\lambda-T(\tau)\,\tau) -{\ds \frac{\widehat{{\bf a}\cdot{\bf b}}}{\widehat{{\bf a}^2}}}\,\delta T\\
 {\ds \frac{1}{L_e^2}} &:={\ds \frac{1}{2L_0(\lambda)^2}+\frac{1}{2L_0(\tau)^2}}.\end{array}\end{equation}$\widehat{{\bf a}^2}$, $\widehat{{\bf b}^2}$ and $\widehat{{\bf a}\cdot {\bf b}}$ are determined by the classical trajectories $\xi_k(\lambda)$. (\ref{it-defn}) results in\begin{equation}\label{ab-defn}\renewcommand{\arraystretch}{2.25} \begin{array}{rl} \widehat{{\bf a}^2} &= {\ds \frac{1}{n}}\, \left({\ds \frac{T(\lambda)}{L_0(\lambda)^2}} +{\ds \frac{T(\tau)}{L_0(\tau)^2}} \right)\\
\widehat{{\bf b}^2} &= {\ds \frac{1}{n}}\, \left({\ds \frac{I(\lambda)-\lambda\dot{I}(\lambda)+\lambda^2 T(\lambda)}{L_0(\lambda)^2}} +{\ds \frac{I(\tau)-\tau\dot{I}(\tau)+\tau^2 T(\tau)}{L_0(\tau)^2}} \right)\\
\widehat{{\bf a}\cdot {\bf b}} &= {\ds \frac{1}{n}}\, \left({\ds \frac{\frac{1}{2} \dot{I}(\lambda)-\lambda T(\lambda)}{L_0(\lambda)^2}} +{\ds \frac{\frac{1}{2} \dot{I}(\tau)-\tau T(\tau)}{L_0(\tau)^2}} \right) \end{array}\end{equation}and $\widehat{{\bf a}}=0$, $\widehat{{\bf b}}=0$ in the selected coordinate frame.
\newline

Substitution and simplification results in the identity\begin{equation}\label{trans-inv}\renewcommand{\arraystretch}{2.75} \begin{array}{rl} \widehat{{\bf a}^2} \widehat{{\bf b}^2} -\widehat{{\bf a}\cdot {\bf b}}^2 &= \left({\ds \frac{T(\lambda)}{nL_0(\lambda)^2}}+{\ds \frac{T(\tau)}{nL_0(\tau)^2}}\right) \left({\ds \frac{I(\lambda)}{nL_0(\lambda)^2}}+{\ds \frac{I(\tau)}{nL_0(\tau)^2}}\right)\\
 &\qquad -\frac{1}{4}\left({\ds \frac{\dot{I}(\lambda)}{nL_0(\lambda)^2}}+{\ds \frac{\dot{I}(\tau)}{nL_0(\tau)^2}}\right)^2 +(\lambda-\tau) {\ds \frac{T(\lambda)}{nL_0(\lambda)^2}}{\ds \frac{\dot{I}(\tau)}{nL_0(\tau)^2}}\\
 &\qquad -(\lambda-\tau) {\ds \frac{T(\tau)}{nL_0(\tau)^2}}{\ds \frac{\dot{I}(\lambda)}{nL_0(\lambda)^2}} +(\lambda-\tau)^2 {\ds \frac{T(\lambda)}{nL_0(\lambda)^2}}{\ds \frac{T(\tau)}{nL_0(\tau)^2}}.\end{array}\end{equation}The $\lambda$ dependence external to the $\xi_k$ and $L_0$ in the magnitude of $C(\lambda,\tau)$ appears as $\lambda-\tau$. 

The demonstration evaluates the connected contribution to $\langle U(\lambda) \underline{s}(\lambda) | \underline{s}(0) \rangle$ and then uses (\ref{lmda-trans}) to evaluate the more general case. The development begins with substitution of the minimum packet functions (\ref{n-minpak}) into the connected VEV (\ref{vev-connctd}) with $\tau=0$.\[\renewcommand{\arraystretch}{1.25} \begin{array}{l} C(\lambda,0) = {\ds \int} d({\bf p})_{2n} \;\delta(\hat{p}_1 \ldots \!+\!\hat{p}_n\!-\!\hat{p}_{n+1} \ldots \!-\!\hat{p}_{2n})\times\\
 \qquad \qquad \qquad L_0(\lambda)^{3n} {\ds \prod_{j=1}^n} e^{i\omega_j \lambda}\, \overline{\tilde{\varphi}({\bf p}_j;\xi_j,\lambda)}\;L_0(0)^{3n} {\ds \prod_{j=n\!+\!1}^{2n}} \tilde{\varphi}({\bf p}_j;\xi_j,0)\end{array}\]with the designation\[\hat{p}_k:=\omega_k,{\bf p}_k\]using (\ref{omega}) for energy-momentum vectors on the positive energy mass shell. The definition (\ref{mapf}) of the $*$-mapped functions, relabeling of the first $n$ momentum summations ${\bf p}_j\mapsto -{\bf p}_j$, the symmetry of the VEV (\ref{vev-connctd}), the UQFT Hamiltonian (\ref{hamil}), and evaluation of the mass shell deltas result in the expression.

The Fourier transforms of the energy-momentum conserving delta generalized function and substitution for the minimum packet functions (\ref{minpak-p}) results in\[C(\lambda,0) = {\ds \int} d({\bf p})_{2n} \; {\ds \int \frac{du}{(2\pi)^4}}\;{\ds \frac{1}{\pi^{3n}}} {\ds \prod_{k=1}^{2n}}L_k^3 e^{is_k\hat{p}_ku} e^{-i s_k\omega_k \tau_k} e^{-\overline{\ell_k^2}({\bf p}_k-{\bf q}_k)^2+is_k ({\bf p}_k-{\bf q}_k)\cdot \xi_k}\]with the shorthand notation\begin{equation}\label{fold}\renewcommand{\arraystretch}{1.25} \begin{array}{ll} s_k=-1, & s_{n+k}=1\\
\tau_k=\lambda, &\tau_{n+k}=0\\
L_k=L_0(\lambda), & L_{n+k}=L_0(0)\\
\ell_k^2=L_0(\lambda)^2-i\frac{1}{2}\lambda_c \lambda,\qquad & \ell_{n+k}^2=L_0(0)^2\\
\xi_j=\xi_j(\lambda), &\xi_{n+j}=\xi_j(0)\\
{\bf q}_j={\bf q}_j(\lambda) & {\bf q}_{n+j}={\bf q}_j(0)\end{array}\end{equation}with $1\leq j\leq n$.

Substitution for $\hat{p}_k$ with the approximation (\ref{nr-p-approx}) for $\omega_k$, the expression (\ref{q-defn}) for ${\bf q}_k$ from $\dot{\xi}_k$, and the nonrelativistic approximation $\lambda_c \omega({\bf q}_k)\approx 1$ from (\ref{nr-approx}) provides that\[\renewcommand{\arraystretch}{1.25} \begin{array}{rl} \hat{p}_ku -\omega_k \tau_k&= \omega_k (u_0-\tau_k) -{\bf p}_k\!\cdot\! {\bf u}\\
 &\approx (\omega({\bf q}_k) + \frac{{\bf q}_k \cdot ({\bf p}_k-{\bf q}_k)}{\omega({\bf q}_k)}+\frac{({\bf p}_k-{\bf q}_k)^2}{2\omega({\bf q}_k)}) (u_0-\tau_k) -{\bf p}_k\!\cdot\! {\bf u}\\
 &\approx (\omega({\bf q}_k) + \dot{\xi}_k\!\cdot\! ({\bf p}_k-{\bf q}_k)+\frac{1}{2}\lambda_c ({\bf p}_k-{\bf q}_k)^2) (u_0-\tau_k)- {\bf p}_k\!\cdot\! {\bf u}.\end{array}\]

For minimum packet functions of sufficient extent for accuracy of the nonrelativistic approximation (\ref{L0-ineq}), the scalar product is readily estimated. Relabeling the summations using ${\bf p}_k':={\bf p}_k-{\bf q}_k$, dropping the prime in the notation, and reorganizing the exponents results in\[C(\lambda,0) = {\ds \int} {\ds \frac{d({\bf p})_{2n}}{\pi^{3n}}} \; {\ds \int \frac{du}{(2\pi)^4}} {\ds \prod_{k=1}^{2n}}L_k^3\, e^{i\omega({\bf q}_k) s_k (u_0-\tau_k)+is_k {\bf q}_k\cdot {\bf u}} e^{is_k {\bf p_k}\cdot (\dot{\xi}_k (u_0-\tau_k)- {\bf u} +\xi_k)}e^{-L_k^2 {\bf p}_k^2}.\]

Designate\begin{equation}\label{deltaq-defn} \delta {\bf q}:= \sum_{j=1}^{2n} s_j {\bf q}_j=\frac{1}{\lambda_c}\left(\sum_{j=1}^n \dot{\xi}_j(0)-\sum_{j=1}^n \dot{\xi}_j(\lambda)\right).\end{equation}To first contributing order for nonrelativistic classical velocities in the approximation (\ref{nr-approx}) for $\omega({\bf q}_k)$, designate\begin{equation}\label{deltat-defn}\renewcommand{\arraystretch}{2.25} \begin{array}{rl} \delta T &:= {\ds \sum_{j=1}^{2n}} s_j \omega({\bf q}_j)\\
 &\approx {\ds \frac{1}{2\lambda_c}}\left({\ds \sum_{j=1}^n} \dot{\xi}_j(0)^2-{\ds \sum_{j=1}^n} \dot{\xi}_j(\lambda)^2\right)\\
 &={\ds \frac{1}{\lambda_c}}(T(0)-T(\lambda))\end{array}\end{equation}and\[\renewcommand{\arraystretch}{2.25} \begin{array}{rl} \phi_T&:= -{\ds \sum_{j=1}^{2n}} s_j \omega({\bf q}_j)\,\tau_j\\
 &\approx {\ds \sum_{j=1}^{n}} \left( {\ds \frac{1}{\lambda_c}}(1+{\ds \frac{\dot{\xi}_j(\lambda)^2}{2}})\lambda \right)\\
 &={\ds \frac{n}{\lambda_c}}\,\lambda + {\ds \frac{1}{\lambda_c}}\, T(\lambda)\lambda.\end{array}\]$\delta {\bf q}$, $\delta T$ and $\phi_T$ are independent of $u$ and $({\bf p})_{2n}$. Conservation of momentum in the classical model provides that $\delta {\bf q}=0$, but for this calculation of the connected contribution to the scalar product (\ref{scalarprod}), more general sets of trajectories $\xi_k(\lambda)$, in particular, scalar products considered in the evaluation of scattering amplitudes, are included. The result after the indicated substitutions is\[C(\lambda,0) = e^{i\phi_T}\; {\ds \int} {\ds \frac{d({\bf p})_{2n}}{\pi^{3n}}} \; {\ds \int \frac{du}{(2\pi)^4}}
 e^{i\,\delta T u_0+i\delta {\bf q}\cdot {\bf u}} \;{\ds \prod_{k=1}^{2n}}\; L_k^3 e^{is_k {\bf p_k}\cdot (\dot{\xi}_k (u_0-\tau_k) - {\bf u} +\xi_k)}e^{-L_k^2{\bf p}_k^2}.\]

The result is a Fourier transform of the product of Gaussian functions $e^{-L_k^2{\bf p}_k^2+i\beta_k\cdot{\bf p}_k}$ and a summation over $u$. From the Gaussian summations (\ref{gauss-sum}) and $s_j^2=1$, the Fourier transforms result in\[\renewcommand{\arraystretch}{1.25} \begin{array}{l} C(\lambda,0) = C_1\; {\ds \int} du \; e^{i\,\delta T u_0+i\delta {\bf q}\cdot {\bf u}}\;{\ds \prod_{k=1}^{2n}}\; e^{-(\dot{\xi}_k (u_0-\tau_k)- {\bf u} +\xi_k)^2/(4L_k^2)}\end{array}\]with\[C_1:=\frac{e^{i\phi_T}}{(2\pi)^4}.\]

From (\ref{fold}) and with $1\leq k \leq n$, the notation is simplified by the designations\begin{equation}\label{vec-defn} \renewcommand{\arraystretch}{1.25} \begin{array}{ll} {\bf a}_k:=\dot{\xi}_k(\lambda),\qquad & 
{\bf a}_{n+k}:=\dot{\xi}_k(0)\\
 {\bf b}_k:= \xi_k(\lambda)-\dot{\xi}_k(\lambda)\,\lambda,\qquad& {\bf b}_{n+k}:= \xi_k(0).\end{array}\end{equation}Weighted means are designated as\begin{equation}\label{mean-defn}\widehat{y}:={\ds \frac{1}{2n}\sum_{j=1}^{2n}}\; \frac{y_j}{L_j^2}\end{equation}and in particular, designate\begin{equation}\label{elle-defn} \frac{1}{L_e^2}:= \widehat{1}=\frac{1}{2n} \sum_{k=1}^{2n}\frac{1}{L_k^2}= \frac{1}{2L_0(\lambda)^2}+\frac{1}{2L_0(0)^2}.\end{equation}$\widehat{y}$ is applied for $y_k ={\bf a}_k^2,{\bf b}_k^2,{\bf a}_k\cdot {\bf b}_k$ and for each of the components of the spatial vectors ${\bf a}_k$ and ${\bf b}_k$. The ${\bf b_k}$ have the units of length, the ${\bf a}_k$ are without units, $L_0(\lambda)$ and $\lambda$ are lengths, and $u=u_0,{\bf u}$ is a spacetime vector of lengths.

Substitution of the definitions (\ref{it-defn}) for $T,I,\dot{I}$ and of ${\bf a}_k,{\bf b}_k$ from (\ref{fold}) in $\widehat{y}$ result in the expressions for $\widehat{{\bf a}^2}$, $\widehat{{\bf b}^2}$ and $\widehat{{\bf a}\cdot {\bf b}}$ in terms of $T(\lambda)$, $I(\lambda)$, $\dot{I}(\lambda)$ and $L_0(\lambda)$ presented in the statement of Lemma 2 as (\ref{ab-defn}). This evaluation applies in a coordinate frame collocated with the classical center of mass for trajectories that exhibit conservation of momentum. In such a frame, $\widehat{\bf a}=\widehat{\bf b}=0$.

Substitution of (\ref{vec-defn}) and (\ref{mean-defn}) in an exponent in the expression for $C(\lambda,0)$ results in\[\renewcommand{\arraystretch}{2.25} \begin{array}{ll} {\ds \sum_{k=1}^{2n}} {\ds \frac{(\dot{\xi}_k (u_0-\tau_k)- {\bf u} +\xi_k)^2}{4L_k^2}} &= {\ds \sum_{k=1}^{2n}} {\ds \frac{({\bf a}_k u_0- {\bf u} +{\bf b}_k)^2}{4L_k^2}}\\
&= {\ds \frac{n}{2}} (\widehat{{\bf a}^2}\, u_0^2+{\ds \frac{{\bf u}^2}{L_e^2}} + \widehat{{\bf b}^2} -2 \widehat{\bf a}\cdot {\bf u}u_0+2\widehat{{\bf a}\cdot{\bf b}}\, u_0 -2 \widehat{\bf b}\cdot {\bf u}).\end{array}\]This substitution, (\ref{ab-defn}) and (\ref{elle-defn}) result in\[C(\lambda,0) = C_1\; {\ds \int} du \; e^{i\,\delta T\, u_0}\; e^{-n({\bf u}^2/L_e^2 +\widehat{{\bf a}^2}\, u_0^2+ \widehat{{\bf b}^2} +2\widehat{{\bf a}\cdot{\bf b}}\, u_0)/2}e^{(i\delta {\bf q}+nu_0\widehat{\bf a}+n\widehat{\bf b})\cdot{\bf u}}.\]The ${\bf u}$ summations result in\[C(\lambda,0) = C_1\,\left({\ds \frac{2\pi L_e^2}{n}}\right)^{\frac{3}{2}}\; {\ds \int} du_0 \; e^{i\,\delta T\, u_0} \;e^{-n(\widehat{{\bf a}^2}\, u_0^2+ \widehat{{\bf b}^2} +2\widehat{{\bf a}\cdot{\bf b}}\, u_0)/2} e^{L_e^2(i\delta {\bf q}+nu_0\widehat{\bf a}+n\widehat{\bf b})^2/2n}.\]Finally, the $u_0$ summation results in\begin{equation}\label{connctd-genrl} C(\lambda,0) = k_C\; e^{-n (\widehat{v}({\bf a}^2)\; \widehat{v}({\bf b}^2)- \widehat{v}({\bf a}\cdot{\bf b})^2)/(2 \widehat{v}({\bf a}^2))}e^{-(\delta T+L_e^2 \delta{\bf q}\cdot\widehat{\bf a})^2/(2n \widehat{v}({\bf a}^2))}e^{-L_e^2\delta{\bf q}^2/(2n)}\end{equation}with\[k_C=C_1 \left({\ds \frac{2\pi L_e^2}{n}}\right)^{\frac{3}{2}} \left({\ds \frac{2\pi}{n\widehat{v}({\bf a}^2)}}\right)^{\frac{1}{2}}\,e^{-i(\delta T+L_e^2\delta{\bf q}\cdot\widehat{\bf a})\,\widehat{v}({\bf a}\cdot{\bf b})/\widehat{v}({\bf a}^2)}e^{iL_e^2\delta{\bf q}\cdot\widehat{\bf b}}\]and\[\widehat{v}({\bf x}\cdot{\bf y}):=\widehat{{\bf x}\cdot{\bf y}}-L_e^2\,\widehat{{\bf x}}\cdot \widehat{{\bf y}}\]for spatial vectors ${\bf x}$ and ${\bf y}$. This is the connected contribution to the scalar product (\ref{scalarprod}) for states labeled by $n$ particle minimum packet functions in the nonrelativistic limit. For trajectories that satisfy Newton's equation, momentum is conserved and as a consequence $\delta{\bf q}=0$ and Poincar\'{e} invariance enables selection of a coordinate frame with $\widehat{\bf a}=\widehat{\bf b}=0$. For these trajectories in the selected coordinate frame, the result simplifies to (\ref{reslt-connctd}).

From\[\widehat{{\bf a}\cdot{\bf b}} =\frac{1}{2n} \sum_{k=1}^{2n} \frac{{\bf a}_k {\bf b}_k}{L_k^2},\]the Cauchy-Schwarz-Bunyakovski inequality provides that\[\widehat{{\bf a}\cdot{\bf b}}^2 \leq \widehat{{\bf a}^2}\; \widehat{{\bf b}^2}\]and the exponential functions in (\ref{reslt-connctd}) are less than unity in magnitude. Also from the Cauchy-Schwarz-Bunyakovski inequality, $\widehat{v}({\bf a}^2)$ and $\widehat{v}({\bf b}^2)$ are both definite in cases of interest, that is, not all ${\bf a}_j= 0$ nor ${\bf b}_j= 0$.

The nonrelativistic approximation to $\exp(i\omega_k \lambda)$ degrades for large periods. With $\epsilon$ the error in the approximation for $\omega_k$, approximation of $\exp(i\omega_k \lambda)$ will have degraded significantly unless $\epsilon \lambda\ll \pi$. From the bound (\ref{taylor}) on the error in the approximation of $\omega_k$,\[|\epsilon| \leq \frac{({\bf p}_k-{\bf q}_k)^4}{2\omega({\bf q}_k)^3} +\frac{2{\bf q}_j^2 \, ({\bf p}_k-{\bf q}_k)^2}{\omega({\bf q}_k)^3} \approx \frac{1}{2} \lambda_c^3 ({\bf p}_k-{\bf q}_k)^4 + 2\lambda_c \dot {\xi}_j^2 ({\bf p}_k-{\bf q}_k)^2\]from (\ref{q-defn}). The significant support of the minimum packet functions (\ref{minpak-p}) is within\[L_0^2 ({\bf p}_k-{\bf q}_k)^2 \leq k_0^2\]for $3\leq k_0 \leq 10$ typically sufficing. For evolution from $\lambda=0$ forward, the approximation will apply for\[\lambda\; ( \frac{\lambda_c^3}{2L_0(\lambda)^4} + 2\dot {\xi}_j^2 \,\frac{\lambda_c}{L_0(\lambda)^2})\ll \pi.\]The bound on $\lambda$ results from the strongest constraint that occurs in the selected center of mass coordinate frame over $1\leq j \leq n$. Then, the nonrelativistic approximation in the evaluation of $\langle U(\lambda) \underline{s}(\lambda)|\underline{s}(0)\rangle$ applies when\begin{equation}\label{lambda-ineq} 0 \leq \lambda \ll\; \frac{\pi}{2\lambda_c}\, \frac{L_0(\lambda)^2}{\lambda_c^2/4L_0(\lambda)^2 + \max_j \dot {\xi}_j^2}.\end{equation}The nonrelativistic approximation degrades for particle-like states, those with small $L_0$. An $L_0(\lambda)\rightarrow \infty$ as $\lambda \rightarrow \infty$ is sufficient to ensure that the approximations apply for large periods. In strictly bound cases, the nonrelativistic approximations fail beyond a limited period.

\subsection{Forward contributions}\label{sec-forw}

The forward contributions $F(\lambda,\tau)$ to the scalar product (\ref{scalarprod}) are evaluated in this section for UQFT VEV and minimum packet functions $\tilde{s}_n((p)_n;\lambda)\in {\cal B}$ centered on classical, $n$ particle trajectories (\ref{n-minpak}).
\newline

\addtocounter{theorems}{1}
{\bf Lemma} \thetheorems: For the forward VEV (\ref{vev-forw}) and minimum packet functions (\ref{n-minpak}), the forward contributions (\ref{S-f-defn}) to the scalar product (\ref{scalarprod}) are\begin{equation}\label{reslt-forw}F(\lambda,\tau) \approx \left({\ds \frac{1}{2\pi}}\right)^\frac{3n}{2}\! \left({\ds \frac{2}{\lambda_c}}\right)^nL_e^{3n}e^{i\theta_F}\, {\ds \sum_{\mathit{pairs}} \prod_{k=1}^n} e^{-L_e^2({\bf q}_k-{\bf q}_{i_k})^2/2} e^{-(\xi_k-\lambda\,\dot{\xi}_k-\xi_{i_k}+\tau\dot{\xi}_{i_k})^2/4L_s^2}e^{i\theta_{ki_k}}\end{equation}in nonrelativistic classical particle instances with\[\renewcommand{\arraystretch}{2.25} \begin{array}{rl} L_s^2 &:=L_0(\lambda)^2+L_0(\tau)^2\\
 \theta_F &:=n\, {\ds \frac{1}{\lambda_c}}\, (\lambda-\tau)+ {\ds \frac{1}{\lambda_c}}\,(T(\lambda)\lambda-T(\tau)\tau)\\
 \theta_{ki_k}&:=({\bf q}_{i_k}-{\bf q}_k)\cdot {\ds \frac{L_0(\lambda)^2\,(\xi_{i_k}-\tau \dot{\xi}_{i_k})+L_0(\tau)^2\,(\xi_k-\lambda\,\dot{\xi}_k)}{L_s^2}}\end{array}\]and $L_e^2$ from (\ref{defn-connctd}).\newline

From (\ref{defn-connctd}), $\theta_C=\theta_F -\delta T \,\widehat{{\bf a}\cdot{\bf b}}/\widehat{{\bf a}^2}$ and $\theta_F$ is an immaterial, overall phase of the ray representing the state.

The demonstration evaluates the forward contribution to $\langle U(\lambda) \underline{s}(\lambda) | \underline{s}(0) \rangle$ and again uses (\ref{lmda-trans}) for the more general case. The development of (\ref{reslt-forw}) is similar to the development of (\ref{reslt-connctd}) and begins with evaluation of the mass shell delta functions after substitution of the minimum packet functions (\ref{n-minpak}) into the forward VEV (\ref{vev-forw}).\[\renewcommand{\arraystretch}{1.25} \begin{array}{rl} F(\lambda,0) &= {\ds \int} d({\bf p})_{2n} \;{\ds \sum_{\mathit{pairs}} \;\prod_{k=1}^n} 2\omega_k \delta({\bf p}_k-{\bf p}_{i_k})\times\\
 &\qquad L_0(\lambda)^{3n}L_0(0)^{3n} {\ds \prod_{j=1}^n} e^{i\omega_j \lambda}\, \overline{\tilde{\varphi}({\bf p}_j;\xi_j,\lambda)} {\ds \prod_{\jmath=n+1}^{2n}} \tilde{\varphi}({\bf p}_\jmath;\xi_\jmath,0)\\
 &= {\ds \int} {\ds \frac{d({\bf p})_{2n}}{\pi^{3n}}} \;{\ds \sum_{\mathit{pairs}} \;\prod_{k=1}^n} 2\omega_k \delta({\bf p}_k-{\bf p}_{i_k})\times\\
 & \qquad {\ds \prod_{j=1}^{2n}}L_j^3 e^{-i s_j\omega_j \tau_j} e^{-\overline{\ell_j^2}({\bf p}_j-{\bf q}_j)^2+is_j ({\bf p}_j-{\bf q}_j)\cdot \xi_j}.\end{array}\]Again, the definition (\ref{mapf}) of the $*$-mapped function, relabeling of the first $n$ momentum summation variables ${\bf p}_j\mapsto -{\bf p}_j$, evaluation of the mass shell deltas, and the notation of (\ref{fold}) simplify the expression. Evaluation of the $n$ momentum delta functions, and the nonrelativistic approximation (\ref{nr-p-approx}) accurate for minimum packet functions of appropriate size (\ref{L0-ineq}) and nonrelativistic relative motion produces\[\renewcommand{\arraystretch}{1.75} \begin{array}{rl} F(\lambda,0) &= e^{i\theta_F}\; {\ds \int} {\ds \frac{d({\bf p})_n}{\pi^{3n}}} \;{\ds \sum_{\mathit{pairs}}}\; {\ds \prod_{k=1}^n} 2\omega_k \times\\
 & \qquad L_k^3 e^{i\dot{\xi}_k\cdot ({\bf p}_k-{\bf q}_k) \lambda} e^{-L_k^2({\bf p}_k-{\bf q}_k)^2-i ({\bf p}_k-{\bf q}_k)\cdot \xi_k}\times\\
 & \qquad L_{i_k}^3 e^{-L_{i_k}^2({\bf p}_k-{\bf q}_{i_k})^2+i ({\bf p}_k-{\bf q}_{i_k})\cdot \xi_{i_k}}.\end{array}\]

The remaining $n$ momentum summations follow from an approximation using the first contribution in the nonrelativistic limit of $\lambda_c \omega_k\approx 1$ from (\ref{nr-approx}) and the Gaussian quadrature\[\renewcommand{\arraystretch}{1.75} \begin{array}{l} L_k^3 L_j^3 {\ds \int} d{\bf p}\; e^{-L_k^2({\bf p}-{\bf q}_k)^2+i({\bf p}-{\bf q}_k)\cdot {\bf h}_k}e^{-L_j^2({\bf p}-{\bf q}_j)^2+i({\bf p}-{\bf q}_j)\cdot {\bf h}_j}\\
 \qquad \qquad \qquad \qquad =\left({\ds \frac{L_e^2 \pi}{2}}\right)^\frac{3}{2} e^{-L_e^2({\bf q}_k-{\bf q}_j)^2/2 -({\bf h}_k-{\bf h}_j)^2/4(L_k^2+L_j^2)} e^{i\theta_{kj}}\end{array}\]with $L_k=L_0(\lambda)$, $L_j=L_0(0)$, ${\bf h}_k:=\xi_k(\lambda)-\lambda \dot{\xi}_k(\lambda)$, ${\bf h}_j:=\xi_{i_k}(0)$, ${\bf q}_j={\bf q}_{i_k}(0)$ and\[\theta_{kj}:=({\bf q}_j(0)-{\bf q}_k(\lambda))\cdot \frac{L_k^2\,{\bf h}_j+L_j^2\,{\bf h}_k}{L_k^2+L_j^2}.\]The indicated substitutions and $L_e^2$ from (\ref{defn-connctd}) result in the desired expression (\ref{reslt-forw}).

The forward contribution to $\langle U(\lambda) \underline{s}(\lambda) | \underline{s}(0) \rangle$ is small in instances with sufficiently distinct incoming from outgoing momenta,\[L_s \|{\bf q}_k(\lambda)-{\bf q}_j(0)\|\gg 1\]with $j,k\in\{1,\ldots n\}$, or for sufficiently great deviations from linear trajectories,\[\|\xi_k(\lambda)-\dot{\xi}_k(\lambda)\lambda-\xi_j(0)\| \gg L_s.\]The close approach of distinct trajectories is precluded by the bounds (\ref{L0-ineq}) on $L_0(\lambda)$.

\subsection{Normalizations and plane wave scattering amplitudes}\label{sec-norm}

The normalizations for UQFT states labeled by minimum packet functions (\ref{n-minpak}) include connected and forward contributions from (\ref{reslt-connctd}) and (\ref{reslt-forw}). (\ref{expand}) from Section \ref{sec-vev} expresses the norm (\ref{norm}) as\[\|\underline{s}(\lambda)\|_{\cal B}^2 =F(\lambda,\lambda)+ c_{2n}C(\lambda,\lambda).\]The scattering cross section is evaluated from a plane wave, large time limit of Born rule transition likelihoods (\ref{likelihood}).

When $\tau=\lambda$, evaluation of the scalar product (\ref{scalarprod}) results in\[\renewcommand{\arraystretch}{1.25} \begin{array}{l} 
 \delta T = \theta_C =\theta_F =0\\
 L_e =L_0(\lambda)\\
 L_s^2 =2L_0(\lambda)^2\end{array}\]from the introductory discussion of Section \ref{sec-eval}, and (\ref{reslt-connctd}) and (\ref{reslt-forw}) from Lemmas 2 and 3. When $\tau=\lambda$,\[\widehat{{\bf a}^2} = {\ds \frac{2T(\lambda)}{nL_0(\lambda)^2}},\qquad 
\widehat{{\bf b}^2} = {\ds \frac{2I(\lambda)}{nL_0(\lambda)^2}},\qquad
\widehat{{\bf a}\cdot {\bf b}} = {\ds \frac{\dot{I}(\lambda)}{nL_0(\lambda)^2}}\]and\[\widehat{{\bf a}^2}\widehat{{\bf b}^2} - \widehat{{\bf a}\cdot {\bf b}}^2 = {\ds \frac{4 I(\lambda)T(\lambda) - \dot{I}(\lambda)^2}{(nL_0(\lambda)^2)^2}}.\]For the connected contributions to the norm,\begin{equation}\label{norm-c}C(\lambda,\lambda) = {\ds \frac{L_0(\lambda)^3}{(2\pi n)^2 \sqrt{\widehat{{\bf a}^2}}}} \; e^{-n (\widehat{{\bf a}^2}\; \widehat{{\bf b}^2}- \widehat{{\bf a}\cdot{\bf b}}^2)/2 \widehat{{\bf a}^2}}\end{equation}and $\widehat{{\bf a}^2}$ is proportional to $L_0^{-2}$.

From Lemma 3, the forward contributions to norms are\begin{equation}\label{norm-f}\renewcommand{\arraystretch}{1.25} \begin{array}{rl} F(\lambda,\lambda) &= k_F {\ds \sum_{\mathit{pairs}} \prod_{k=1}^n} e^{-L_0^2({\bf q}_k-{\bf q}_{i_k})^2/2} e^{-(\xi_k-\xi_{i_k})^2/8L_0^2}e^{i\theta_{ki_k}}\\
 &\approx k_F\end{array}\end{equation}with\[\renewcommand{\arraystretch}{2.25} \begin{array}{rl} k_F &:=\left({\ds \frac{2}{\lambda_c}}\right)^n \left({\ds \frac{L_0}{\sqrt{2\pi}}}\right)^{3n}\\
 \theta_{ki_k}&:={\ds \frac{1}{2}} ({\bf q}_{i_k}-{\bf q}_k)\cdot (\xi_{i_k}+\xi_k).\end{array}\]From (\ref{fold}) and when $\tau=\lambda$, $\xi_{n+k}=\xi_k$. The oddness of the phase,\[\theta_{i_k k}=-\theta_{k i_k},\]and the symmetry of the VEV (\ref{vev-forw}) under transposition of arguments verifies that the forward contribution to the norm is real and nonnegative. The nonrelativistic classical particle conditions (\ref{L0-ineq}) provide that with sufficiently diverse trajectories, the summation labeled {\em pairs} is dominated by the term with $i_k-n=k$ for each $k\in \{1,\ldots n\}$. $(\xi_j-\xi_k)^2 \geq r_a^2$ with $r_a$ the closest approach of two distinct trajectories, and (\ref{L0-ineq}) provides that $L_0\ll r_a$. As a consequence, $\exp(-(\xi_j(\lambda)-\xi_k(\lambda))^2/8L_0^2)\ll 1$ except when $j=k$.

From the connected (\ref{norm-c}) and forward (\ref{norm-f}) contributions to norms, $F(\lambda,\lambda)$ scales as $L_0^{3n}$ and the $n\geq 2$ connected contribution $c_{2n}C(\lambda,\lambda)$ scales as $L_0^4$. As a consequence, for sufficiently small $L_0$, the connected contribution dominates the norm and for sufficiently large $L_0$, the forward contribution dominates the norm. Small $L_0$ are associated with particle-like cases, and $L_0$ increases without bound in plane wave cases. The $L_0$ dependence of the partially factored contributions are bounded between these two extremes.

Limits of the minimum packet functions (\ref{n-minpak}) with $L_0(\lambda)$ increasing without bound describe plane waves that label states that diverge from the Hilbert space. In plane wave limits, the minimum packet functions coincide with functions used by Lehmann, Symanzik, and Zimmermann (LSZ) [\ref{bogo}] in their studies of scattering in QFT and used in [\ref{intro},\ref{gej05},\ref{feymns}] to evaluate UQFT scattering amplitudes. In this divergent case, linear combinations of the limits using appropriate measures on subsets of states determine projection operators. Plane waves lack spatial features except for a time varying phase determined by the momenta $({\bf q})_n$ and, as a consequence, up to an overall phase it suffices to label the plane wave limits of minimum packet states $|\underline{s}(\lambda)\rangle$ by the momenta $({\bf q})_n$ determined from the classical trajectories. In this case, the projection operator in the Born's rule likelihood (\ref{likelihood}) is expressed\begin{equation}\label{norm-wave}P_f= \left(\frac{\lambda_c}{2}\right)^n \int_{W_\sigma} d({\bf q})_n\;|U(\lambda)\underline{s}(\lambda)\rangle \langle U(\lambda)\underline{s}(\lambda)|.\end{equation}$W_\sigma$ is a Lebesgue measurable subset of ${\bf R}^{3n}$ and the $|U(\lambda)\underline{s}(\lambda)\rangle$ are parameterized by $({\bf q})_n$. This normalization results from the idempotence of $P_f$ in the plane wave limit.

With the momentum parameters ${\bf q}_k$ of states in the summations explicitly displayed,\[\renewcommand{\arraystretch}{2.25}\begin{array}{rl} \langle \underline{s}(\lambda;({\bf q}')_n)|\underline{s}(\lambda;({\bf q})_n)\rangle &\approx k_F {\ds \sum_{\mathit{pairs}}\prod_{k=1}^n} e^{-L_0^2({\bf q}'_k-{\bf q}_{i_k})^2/2}\\
 &\approx k_F \left({\ds \frac{\sqrt{2\pi}}{L_0}}\right)^{3n} {\ds \sum_{\mathit{pairs}}\prod_{k=1}^n} \delta({\bf q}_k-{\bf q'}_k)\\
 &\approx {\ds \left(\frac{2}{\lambda_c}\right)^n} {\ds \sum_{\mathit{pairs}}\prod_{k=1}^n} \delta({\bf q}_k-{\bf q'}_k) \end{array}\]in the large $L_0(\lambda)$ limit of the scalar product (\ref{reslt-forw}) using the delta sequences (\ref{delta-seq}). The dominance of the forward contribution to the scalar product in the plane wave limit verifies that $P_f$ is a projection.\[ \renewcommand{\arraystretch}{1.75}\begin{array}{rl} P_f^2 &= {\ds \left(\frac{\lambda_c}{2}\right)^{2n}} {\ds \iint} d({\bf q})_n d({\bf q}')_n\; 
\langle \underline{s}(\lambda;({\bf q}')_n) |\underline{s}(\lambda;({\bf q})_n)\rangle\; | U(\lambda) \underline{s}(\lambda;({\bf q}')_n) \rangle\,\langle U(\lambda) \underline{s}(\lambda;({\bf q})_n)|\\
 &\approx {\ds \left(\frac{\lambda_c}{2}\right)^n} {\ds \int}_{W_\sigma} d({\bf q})_n {\ds \int}_{W_\sigma}d({\bf q}')_n\; 
{\ds \sum_{\mathit{pairs}} \prod_{k=1}^n}\delta({\bf q}_k-{\bf q}_k')\; | U(\lambda) \underline{s}(\lambda;({\bf q}')_n) \rangle\,\langle U(\lambda) \underline{s}(\lambda;({\bf q})_n) | \\
 &= P_f.\end{array}\]assuming that the subset $W_\sigma$ consists of small neighborhoods of distinct momenta, ${\bf q}_j\neq {\bf q}_k$ when $({\bf q})_n\in W_\sigma$.

The elastic scattering cross sections derive from the large time, plane wave limits of the Born's rule transition likelihoods with $n=2$.\[\int_{W_\sigma}d\mu_s\;{\cal I}(\lambda,-\lambda)^2=\mbox{Trace}(P_f | U(-\lambda) \underline{u}_s(-\lambda) \rangle\,\langle U(-\lambda) \underline{u}_s(-\lambda) |)\]with the projection operator $P_f$ (\ref{norm-wave}) described by outgoing momenta near ${\bf q}_1,{\bf q}_2$, and the state density matrix is a unit trace projection onto a pure state with momentum support dominantly near ${\bf q}_3,{\bf q}_4$. The incoming state is a temporal translation of the normalized, nonrelativistic classical particle state $|\underline{s}(\lambda) \rangle $ from (\ref{n-minpak}). The likelihood of a transition to any particular momentum vanishes in a plane wave limit but the likelihood of transition into a Lebesgue measurable subset of momenta $W_\sigma$ is finite. The differential likelihood ${\cal I}(\lambda,-\lambda)^2$ is the likelihood that initial states that are nearly plane waves described by incoming momenta ${\bf q}_3,{\bf q}_4$ transition to one of a subset of outgoing energy-momentum vectors near ${\bf q}_1,{\bf q}_2\in W_\sigma =d({\bf q})_2$. All four energy-momenta $(q)_4$ are on the mass $m$ mass shell. The non-forward, large time difference, plane wave limits of the scalar products of interest are\[\renewcommand{\arraystretch}{1.25}\begin{array}{rl} {\ds \lim_{\lambda \rightarrow \infty} \lim_{L_0 \rightarrow \infty}} |\langle U(\lambda)\underline{s}(\lambda) |U(-\lambda)\underline{s}(-\lambda)\rangle| &={\ds \lim_{\lambda \rightarrow \infty} \lim_{L_0 \rightarrow \infty}} c_4 \,|C(\lambda,-\lambda)|\\
 &={\ds \lim_{L_0 \rightarrow \infty}} c_4 \left({\ds \frac{L_e}{2\sqrt{\pi}}}\right)^3 {\ds \frac{1}{2\sqrt{\pi \widehat{v}({\bf a}^2)}}} e^{-(\delta T)^2/(4 \widehat{v}({\bf a}^2))}e^{-L_e^2\delta{\bf q}^2/4}\\
 &\approx c_4 \delta(q_\mathit{dif}).\end{array}\]This result follows from the intermediate result (\ref{connctd-genrl}) from Lemma 2 for general $({\bf q})_n$, and the delta sequences (\ref{delta-seq}) with $\delta {\bf q}$ from (\ref{deltaq-defn}), $\delta T$ from (\ref{deltat-defn}) and for $n=2$.\[q_\mathit{dif}:=q_1+q_2-q_3-q_4\]for Lorentz vectors $q_k$ and the likelihood of a transition vanishes unless energy and momentum are conserved. For a non-forward selection of momenta, the forward contributions are negligible, the connected contribution (\ref{connctd-genrl}) has $\lambda=-\tau$ and $L_0(\tau)=L_0(\lambda)\rightarrow \infty$. From (\ref{reslt-connctd}), $L_e^2=L_0(\lambda)^2$, and from (\ref{ab-defn}), $\widehat{\bf a}$, $\widehat{{\bf b}^2}$, and $(\widehat{{\bf a}\cdot{\bf b}})^2/\widehat{{\bf a}^2}$ become negligible as $L_0(\lambda)$ grows without bound. The plane wave limit precedes the large time limit.

The product of the differential likelihood ${\cal I}(\lambda,-\lambda)^2$ with the flux corrected interaction area $A=V/(u_\alpha T)$ and the identification\[\delta^2(q_\mathit{dif})=\frac{V}{(2\pi)^3}\frac{T}{2\pi} \;\delta(q_\mathit{dif})\]from box normalization [\ref{weinberg}] results in the differential cross section. The volume $V=L_0^3 \sqrt{(\pi/2)^3}$ and duration $T=\sqrt{\pi/(2\widehat{v}({\bf a}^2))}$ both diverge in the plane wave limit. Summed over all ${\bf q}_2$ and all magnitudes $\|{\bf q}_1\|$, the resulting differential cross section is the elastic scattering cross section into a solid angle $d\Omega$. In a (barycentric) reference frame collocated with the classical center of mass, the indicated substitutions result in\[\renewcommand{\arraystretch}{2.25}\begin{array}{rl} {\ds \frac{d\sigma}{d\Omega}}&= \left({\ds \frac{\lambda_c}{2}}\right)^2 {\ds \int d{\bf q}_2 \int_0^\infty} d\|{\bf q}_1\| \;\|{\bf q}_1\|^2 {\ds \frac{V}{u_\alpha T}} {\ds \frac{|\langle U(\lambda)\underline{s}(\lambda)|U(-\lambda)\underline{s}(-\lambda)\rangle|^2}{\|\underline{s}(-\lambda)\|^2}} \\
 &= {\ds \int d{\bf q}_2 \int_0^\infty} d\|{\bf q}_1\| \;\|{\bf q}_1\|^2 {\ds \frac{(c_4)^2}{2\pi} \left(\frac{\lambda_c}{2}\right)^4 \left(\frac{\pi}{2}\right)^3 \frac{\omega({\bf q}_3)}{2\|{\bf q}_3\|}}\; \delta(q_\mathit{dif})\end{array}\]from\[u_\alpha:=\frac{\sqrt{(q_3q_4)^2-m^4}}{E_3E_4} = \frac{2\|{\bf q}_3\|}{\omega({\bf q}_3)}\]in this case of equal mass particles. The result is\[\frac{d\sigma}{d\Omega}\approx {\ds \frac{(c_4)^2}{8\pi} \left(\frac{\lambda_c}{2}\right)^2 \left(\frac{\pi}{2}\right)^3}\]to first contributing order in the nonrelativistic approximation and has a plane wave limit. Due to the exchange term for indistinguishable particles [\ref{rgnewton}], this nonrelativistic approximation to the cross section can be associated with many equivalent potentials but is perhaps most naturally associated with $V(r)=c \delta(r)$. Relativistic instances are included in [\ref{gej05},\ref{feymns}].
 
This differential cross section $d\sigma/d\Omega$ is independent from scattering angle in the nonrelativistic approximation and not consistent with the Mott cross section for a $1/r$ potential despite the association of the nonrelativistic classical particle trajectories with $1/r$ potentials.
\section{The classical trajectories of a UQFT}\label{sec-corrsp}

In this section, examples of trajectories that optimize the scalar products (\ref{scalarprod}) are studied to associate nonrelativistic classical particles with the UQFT realization of relativistic quantum physics. The optimization minimizes the error in the approximation (\ref{idea}). Maximization of the Born's rule transition likelihood (\ref{likelihood}) minimizes the distinction between temporal translation of states and translation of the dominant regions of support along classical trajectories $\xi_k(\tau)$. The set of $n$ trajectories $\xi_k(\lambda)$ that maximize the value of the likelihood (\ref{likelihood}) are the nonrelativistic classical particle approximations derived from the UQFT. The trajectories are defined in a configuration space established to achieve simple descriptions of motion.

\subsection{Two body trajectories}\label{sec-twop}

Two body problems provide explicit examples. The classical two body problem is soluble and for the $1/r$ potential, the trajectories are expressed in elementary functions.

With two particles, $n=2$ and in coordinates collocated with the center of mass, $\xi_2=-\xi_1$. Newton's equation provides that the motion of the two particles is executed within a plane for forces that depend only on the distance between particles. Poincar\'{e} invariance is exploited to select a frame with\begin{equation}\label{twop-coord}{\bf x}:=\xi_1-\xi_2=2\xi_1:=\left(\begin{array}{c}r \cos\theta\\ r\sin\theta\\0\end{array}\right)\end{equation}with $r:=\|\xi_1-\xi_2\|$, the distance between the two particles, and in the two particle case, $r_a(\lambda)=r(\lambda)$. The angular momentum and classical energy are constants of the motion,\[\begin{array}{rl} L &:= r^2 \dot{\theta}\\
e_C &= {\ds \frac{\dot{r}^2}{4}}+{\ds \frac{L^2}{4r^2}}+V(r).\end{array}\]The quantities used in the evaluation of the connected contributions to the scalar product include\[\renewcommand{\arraystretch}{2.25} \begin{array}{rl} I&={\ds \frac{1}{2}}\,(\xi_1^2+\xi_2^2)={\ds \frac{r^2}{4}}\\
 \dot{I}&={\ds \frac{r\dot{r}}{2}}\\
 T&={\ds \frac{1}{2}}\,(\dot{\xi}_1^2+\dot{\xi}_2^2)={\ds \frac{\dot{r}^2}{4}}+{\ds \frac{L^2}{4r^2}}\end{array}\]from (\ref{it-defn}).

For the $1/r$ potential,\[V=-\frac{g}{r},\]Newton's equation in Jacobi coordinates results in\[\ddot{\bf x}=-\frac{2g}{r^3}\,{\bf x}\]with ${\bf x}$ from (\ref{twop-coord}) and the factor of two from the reduced mass of two equal mass $m$ particles. The solution for the trajectories $\xi_1,\xi_2$ is conveniently parameterized by $\theta$ from (\ref{twop-coord}) and solution of Newton's equation results in\[r(\theta) ={\ds \frac{L^2/2g}{1-\epsilon_r\,\cos \theta}}\]and\[\lambda(\theta)=\frac{1}{L} \int_{\theta_0}^\theta d\phi\; r(\phi)^2\]with $\lambda(\theta_0)=0$ and\[\epsilon_r:= \sqrt{1+e_C L^2/g^2}.\]For $e_C>0$, the solution diverges when $\epsilon_r\cos\theta =1$ and $\theta$ is constrained to the interval $(\beta, 2\pi \!-\! \beta)$ with $\beta=\cos^{-1}(\epsilon_r^{-1})$. For a real solution,\[e_C\geq -\frac{g^2}{L^2}\]and $e_C= -g^2/L^2$ are circular orbits. For circular orbits, the constant $r$ and the classical coupling constant $g$ can be considered as independent variables with $e_C$ and $L$ determined.\[e_C = -{\ds \frac{g}{2r}},\qquad \qquad L= \sqrt{2gr},\qquad \qquad \lambda(\theta) ={\ds \frac{r^\frac{3}{2}}{\sqrt{2g}}}\;\theta\]for $\lambda(0)=0$. To satisfy a nonrelativistic velocity in this case of two particles in circular orbit,\[(r\dot{\theta})^2 = \frac{2g}{r}\ll 1\]requires that $g\ll r$. For gravity, the length $g=Gm/c^2$ in the units of this note. For a mass of one a.m.u., $g$ is $1.3\times10^{-54}$ m.

The likelihoods can be expressed in terms of four lengths, $\lambda_c,r,g$ and $L_0$, and the interaction strength coefficient $c_4$. Substitution for $I,\dot{I},T$ into the definitions for $\widehat{{\bf a}^2}$, $\widehat{{\bf b}^2}$, $\widehat{{\bf a}\cdot{\bf b}}$ from (\ref{ab-defn}) and evaluation of (\ref{trans-inv}) for circular orbits result in\[\renewcommand{\arraystretch}{2.25} \begin{array}{rl} \widehat{{\bf a}^2} &= {\ds \frac{T(0)}{L_0(0)^2}}={\ds \frac{L^2}{4r^2}\frac{1}{L_0^2}}={\ds \frac{g}{2L_0^2 r}}\\
\widehat{{\bf b}^2} &= {\ds \frac{I(0)+\frac{1}{2} \lambda^2 T(0)}{L_0(0)^2}}={\ds \frac{r^2}{4L_0^2}}+{\ds \frac{g\lambda^2}{4L_0^2 r}}\\
\widehat{{\bf a}\cdot {\bf b}} &= -{\ds \frac{\frac{1}{2} \lambda T(0)}{L_0(0)^2}}=-{\ds \frac{g\lambda}{4L_0^2 r}}\\
{\ds \frac{\widehat{{\bf a}^2} \widehat{{\bf b}^2} - \widehat{{\bf a}\cdot {\bf b}}^2}{\widehat{{\bf a}^2}}} &= {\ds \frac{I(0)}{L_0(0)^2}}+{\ds \frac{\lambda^2\, T(0)}{4\, L_0(0)^2}}\\
 &={\ds \frac{r^2+\lambda^2 \,L^2/4 r^2}{4L_0(0)^2}}={\ds \frac{2r^3+\lambda^2 g}{8L_0^2 r}}.\end{array}\]For circular orbits, $T(\lambda)=T(0)$, $I(\lambda)=I(0)$ and $\dot{I}=0$. Substitution in (\ref{reslt-connctd}) from Lemma 2 produces\[C(\lambda,0) = k_C \; e^{- (\widehat{{\bf a}^2}\; \widehat{{\bf b}^2}- \widehat{{\bf a}\cdot{\bf b}}^2)/ \widehat{{\bf a}^2}}=k_C\; e^{-(2r^3+\lambda^2 g)/8L_0^2 r}\]with\[k_C = {\ds \frac{L_0^3 \, e^{i\theta_C}}{(4\pi)^2 \sqrt{\widehat{{\bf a}^2}}}}={\ds \frac{L_0^4\, e^{i\theta_C}}{8\pi^2}}\sqrt{\frac{r}{2g}}\]from $\delta T=0$ in this circular orbit case. Noting that $C(\lambda,\lambda)=C(0,0)$ results in all the connected contributions of interest.

The forward contribution to the scalar product is (\ref{reslt-forw}) from Lemma 3. The identical factors in each term result in\[\renewcommand{\arraystretch}{1.75} \begin{array}{rl} F(\lambda,\tau) &= k_F {\ds \sum_{\mathit{pairs}} \prod_{k=1}^n} e^{-L_e^2({\bf q}_k-{\bf q}_{i_k})^2/2} e^{-(\xi_k-\lambda\,\dot{\xi}_k-\xi_{i_k}+\tau\,\dot{\xi}_{i_k})^2/4L_s^2}e^{i\theta_{ki_k}}\\
 &= k_F e^{-L_0^2({\bf q}_1-{\bf q}_1')^2} e^{-(\xi_1-\lambda\dot{\xi}_1 -\xi'_1 +\tau\dot{\xi}'_1)^2/4L_0^2}\\
 &\qquad + k_F e^{-L_0^2({\bf q}_1+{\bf q}_1')^2} e^{-(\xi_1-\lambda\dot{\xi}_1 +\xi'_1 -\tau\dot{\xi}'_1)^2/4L_0^2}\end{array}\]with the abbreviated notation $\xi_1:=\xi_1(\lambda)$, $\xi'_1:=\xi_1(\tau)$, ${\bf q}_1:={\bf q}_1(\lambda)$, ${\bf q}'_1:={\bf q}_1(\tau)$, and noting that $\theta_F=\theta_C$, $L_s^2=2L_0(0)^2$, $L_e^2=L_0(0)^2$, and\[k_F =\frac{4L_0^6}{\pi^3 \lambda_c^2}\; e^{i\theta_C}\]for this $n=2$, circular orbit case. In this case, $\theta_{11}=\theta_{22}=0$, $\theta_{12}=-\theta_{21}$ and both terms are real except for the common complex unimodular factor in $k_F$ and $k_C$. From the expression (\ref{twop-coord}) for $\xi_1$ and with $\theta := \theta(\lambda)$ and $\theta':=\theta(\tau)$,\[\renewcommand{\arraystretch}{2.25} \begin{array}{l} (\xi_k(\lambda)-\lambda \dot{\xi}_k(\lambda)-\xi_{i_k}(\tau)+\tau \dot{\xi}_{i_k}(\tau))^2 = (\xi_1(\lambda)-\lambda \dot{\xi}_1(\lambda)\pm(\xi_1(\tau)- \tau \dot{\xi}_1(\tau)))^2\\
 \qquad= ({\ds \frac{r}{2}}\cos\theta +\lambda {\ds \frac{L}{2r}}\sin\theta \pm ({\ds \frac{r}{2}}\cos\theta' +\lambda {\ds \frac{L}{2r}}\sin\theta'))^2\\
 \qquad \qquad \qquad +({\ds \frac{r}{2}}\sin\theta -\lambda {\ds \frac{L}{2r}}\cos\theta \pm ({\ds \frac{r}{2}}\sin\theta' -\lambda {\ds \frac{L}{2r}}\cos\theta'))^2\\
 \qquad= {\ds \frac{r^2}{2}}\, (1\pm \cos(\theta\!-\!\theta'))+{\ds \frac{g}{2r}}(\lambda^2+\tau^2 \pm 2\lambda \tau \cos(\theta\!-\!\theta'))\pm {\ds \sqrt{\frac{gr}{2}}}\; (\lambda-\tau) \sin(\theta-\theta').\end{array}\]This result and (\ref{q-defn}) provides that\[\renewcommand{\arraystretch}{2.25} \begin{array}{rl} ({\bf q}_k(\lambda)-{\bf q}_{i_k}(\tau))^2 &= {\ds \frac{1}{\lambda_c^2}} (\dot{\xi}_1(\lambda)\pm \dot{\xi}_1(\tau))^2\\
 &= {\ds \frac{g}{\lambda_c^2 r}} (1\pm \cos (\theta\!-\!\theta')).\end{array}\]From $\lambda=\sqrt{r^3/2g}\; \theta$ for the circular orbits of a $1/r$ potential, substitution yields\[\renewcommand{\arraystretch}{1.75} \begin{array}{rl} F(\lambda,0) &= k_F\, e^{-a_0\,(1+\frac{1}{2}\theta^2-\cos\theta-\theta\sin\theta) -a_1\,(1-\cos\theta)}+k_F\, e^{-a_0\,(1+\frac{1}{2}\theta^2+\cos\theta+\theta\sin\theta) -a_1\,(1+\cos\theta)}\\
 C(\lambda,0) &= k_C\; e^{-a_0\,(2+\frac{1}{2}\theta^2)}.\end{array}\]with two constants\begin{equation}\label{two-const}a_0 := {\ds \frac{r^2}{8 L_0^2}},\qquad \qquad \qquad \qquad
 a_1 := {\ds \frac{L_0^2g}{\lambda_c^2 r}}.\end{equation}Inspection provides that $F(0,0)=F(\lambda,\lambda)$ and all the forward terms of interest result.

In this case of two particles in circular orbit for a $-g/r$ potential, the amplitude (\ref{i-defn}) that in this case is the square root of the transition likelihood is\[{\cal I}(\lambda,0) = {\ds \frac{|F(\lambda,0)+c_4C(\lambda,0)|}{F(0,0)+c_4C(0,0)}}\]and the indicated substitutions result in\[{\cal I} = \frac{e^{-a_0\,(\frac{1}{2}\theta^2\!-\cos\theta\!-\theta\sin\theta) -a_1\,(1\!-\cos\theta)}+e^{-a_0\,(\frac{1}{2}\theta^2\!+\cos\theta\!+\theta\sin\theta) -a_1\,(1\!+\cos\theta)}+ c_R\,e^{-a_0\,(1\!+\frac{1}{2}\theta^2)}}{e^{a_0}+e^{-a_0-2a_1}+ c_R\, e^{-a_0}}\]with\begin{equation}\label{kr}k_R:={\ds \frac{\pi c_4 \lambda_c}{32\sqrt{2} L_0}}\mbox{, and}\qquad c_R:=\frac{k_R}{\sqrt{a_1}}.\end{equation}

For small $\theta$, the likelihoods satisfy the coplanar propagation bound (\ref{near-cop}) in this two body circular orbit case. For sufficiently small $\theta$, the likelihood is approximated by\begin{equation}\label{circ-reslt}{\cal I}(\lambda,0) \approx 1-\frac{\frac{a_1}{2}+(a_0 -\frac{a_1}{2})e^{-2a_0-2a_1} + c_R\,\frac{a_0}{2} \; e^{-2a_0}}{1+e^{-2a_0-2a_1}+ c_R\, e^{-2a_0}}\;\theta^2 \end{equation}and from $a_0,a_1>0$, the likelihood satisfies the coplanar propagation bound (\ref{near-cop}). From the indicated $a>0$, $\theta=\theta_1+\theta_2$ with $\theta_1,\theta_2>0$,\[{\cal I}(\lambda,0)\approx 1- a\, (\theta_1+\theta_2)^2 \leq 1-a\, (\theta_1^2+\theta_2^2) \approx {\cal I}(\lambda_1,0){\cal I}(\lambda_2,\lambda_1).\]For greater $\lambda$ in this same case, validity of the coplanar propagation bound is conditional and, for example, for $\theta_1=\theta_2$, the coplanar propagation bound remains valid up to $\theta=4$ and to likelihoods ${\cal I}(\lambda,0)$ smaller that $10^{-3}$ when $a_1<a_0$ and $c_R<1$. More generally, the coplanar propagation bound applies in instances with dominance of the temporal dependence by $\exp(-a (\lambda_1+\lambda_2)^2)$. $\exp(-a (\lambda_1+\lambda_2)^2)< \exp(-a \lambda_1^2)\exp(-a \lambda_2^2)$ when $a,\lambda_1,\lambda_2 >0$.

Satisfaction of the coplanar propagation bound for this two body circular orbit case and Lemma 1 provide that the likelihoods of the trajectories of $1/r$ potentials have the greatest upper bound. The appropriate classical potential for circular orbits is $-g/r$ but the potential strength $g$ has not been determined. The interaction strength $g$ is determined to provide estimated likelihoods that persist near unity for the greatest period. Displayed in the evaluation (\ref{circ-reslt}) of the amplitude for this circular orbit case, ${\cal I}(0,0)=1$ and $d{\cal I}/d\lambda=0$ with ${\cal I}={\cal I}(\lambda,0)$ evaluated at $\lambda=0$. From the linear relation between $\lambda$ and $\theta$, the $g$ that provides the least negative value,\[\left.\frac{d^2{\cal I}}{d\theta^2}\right|_{\theta=0}=\max,\]determines the likelihoods that persist near unity for the greatest period. From (\ref{circ-reslt}),\[\left. \frac{d^2{\cal I}}{d\theta^2}\right|_{\theta=0}=\frac{-a_1 -(2a_0-a_1)e^{-2a_0 -2a_1}-c_R\, a_0\; e^{-2a_0}}{1+e^{-2a_0-2a_1}+ c_R\, e^{-2a_0}}.\]The nonrelativistic classical particle condition (\ref{L0-ineq}) provides that $r \gg L_0$ and then (\ref{two-const}) results in $a_0\gg 1$. With neglect of contributions from $\exp(-2a_0-2a_1)$ with respect to unity, this simplifies to\[\left. \frac{d^2{\cal I}}{d\theta^2}\right|_{\theta=0}\approx \frac{-a_1 -2a_0 e^{-2a_0 -2a_1}-c_R\, a_0\; e^{-2a_0}}{1+ c_R\, e^{-2a_0}}.\]

Extrema occur at zeroes of the first derivative with respect to $g$, with the result that the likelihood is maximized for\[\renewcommand{\arraystretch}{2.25} \begin{array}{l} \left((-1 +4a_0e^{-2a_0-2a_1}+{\ds \frac{c_R\,a_0e^{-2a_0}}{2a_1}})(1+ c_R\, e^{-2a_0}) \right.\\
 \qquad\qquad \left. +(-a_1 -2a_0 e^{-2a_0 -2a_1}-c_R\, a_0\; e^{-2a_0}){\ds \frac{c_R\, e^{-2a_0}}{2a_1}} \right) {\ds \frac{da_1}{dg}} = 0\end{array}\]that results from\[\frac{dc_R}{da_1}=-\frac{c_R}{2a_1}\]using (\ref{kr}) and that $(1+ c_R\, e^{-2a_0})\neq 0$. This simplifies to\[{\ds \frac{k_Ra_0e^{-2a_0}}{2\sqrt{a_1^3}}}- {\ds \frac{3k_Re^{-2a_0}}{2\sqrt{a_1}}} -1 =0\]with neglect of contributions from $2e^{-2a_0-2a_1}$ and $4a_0 e^{-2a_0-2a_1}$ with respect to unity from $a_0\gg 1$. This is a cubic equation for $1/\sqrt{a_1}$ with no solution when $k_R= 0$. With designations\begin{equation}\label{cubic}x:=\frac{1}{\sqrt{a_1}},\qquad a:=-\frac{3}{a_0},\mbox{ and} \qquad b:=-\frac{2}{k_Ra_0 e^{-2a_0}},\end{equation}the real solution to $x^3+ax+b=0$ is\[x=A+B\]with\[A^3=\frac{ e^{2a_0}}{k_Ra_0} +\sqrt{\frac{e^{4a_0}}{(k_Ra_0)^2}-\frac{1}{(a_0)^3}}\;,\mbox{ and}\qquad B^3=\frac{ e^{2a_0}}{k_Ra_0} -\sqrt{\frac{e^{4a_0}}{(k_Ra_0)^2}-\frac{1}{(a_0)^3}}.\]In instances with sufficiently weak coupling that $a_0e^{4a_0}\gg k_R^2$,\[x\approx \left(\frac{ 2e^{2a_0}}{k_Ra_0}\right)^\frac{1}{3}.\]The strength $g$ of the classical potential is expressed in terms of $c_4$ with substitution for $a_0$ and $a_1$ from (\ref{two-const}), for $k_R$ from (\ref{kr}) and for $x$ from (\ref{cubic}). The result is\[g =\left( \frac{\lambda_c^3 r^\frac{3}{2}k_R a_0 e^{-2a_0}}{2L_0^3} \right)^\frac{2}{3}=\left( \frac{\pi c_4 \lambda_c^4 r^\frac{7}{2}e^{-\frac{r^2}{4L_0^2}}}{512\sqrt{2} L_0^6} \right)^\frac{2}{3}.\]

In this circular orbit case, from (\ref{L0-form}), $L_0=L_0(T,V)=L_0(g/2r,-g/r)$ is constant with variations of $\lambda$ but varies with $r$. The selection of $L_0$ as the solution to\begin{equation}\label{lambda-r} L_0^6 = \beta_0 r^\frac{7}{2} e^{-\frac{r^2}{4L_0^2}}\end{equation}results in a strength for the classical potential\[g= \left(\frac{\pi \lambda_c^4}{512\sqrt{2}\, \beta_0}\, c_4\right)^\frac{2}{3}\]that is independent of the orbit diameter $r$. $\beta_0$ is a physical constant associated with this particular UQFT and $\beta_0$ may be a function of mass $m$ but is independent of $r$.

There are real solutions to (\ref{lambda-r}) only if $r$ is less than an upper bound, roughly\[r \lesssim \left(\frac{12}{e}\right)^\frac{6}{5}\;\beta_0^\frac{2}{5}.\]Quantum corrections to the trajectories with respect to the trajectories of a $1/r$ potential occur at larger radii orbits for this selected UQFT.

As a consequence of the limited duration of the applicability of the nonrelativistic approximations (\ref{lambda-ineq}), better methods are required to examine bound states. The approximations and the upper bound on $L_0(\lambda)\ll r$ from the nonrelativistic classical particle bounds (\ref{L0-ineq}) do not enable extension of this analysis of two body orbits to times $\lambda \rightarrow \infty$ unless $r \rightarrow \infty$ and then deviations from a $-g/r$ potential occur. There is a finite likelihood that any state labeled by Gaussian minimum packet functions escapes the bound state due to the small but finite support at large energies. The generalized eigenstates of the UQFT Hamiltonian are plane waves. However, the association of nonrelativistic classical trajectories with the UQFT and Ehrenfest's theorem associates UQFT with an interaction Hamiltonian of a nonrelativistic quantum theory that includes bound eigenstates. In this nonrelativistic, finite duration approximation, conventional bound states are associated with UQFT by the nonrelativistic, classical trajectories in common.

This two body circular orbit instance displays that the properties of the classical limit determines $L_0(\lambda)$. $L_0(\lambda)$ provides the freedom to match the classical limits of UQFT with observed classical limits. The result is that UQFT VEV are compatible with appropriate classical limits over interesting ranges of parameters.

\subsection{Newton's equations from a UQFT}\label{sec-newteq}

Derivation of classical trajectories associated with the selected UQFT simplifies significantly when particle-like initial states transition to wave-like states with\begin{equation}\label{growth}\frac{L_0(\lambda)}{\lambda} \rightarrow \infty\end{equation}as $\lambda$ grows without bound. In this instance, and with selection of an appropriate dynamic model for the momentum spread length $L_0(\lambda)$, Newton's equations appear naturally in the optimization of the likelihoods (\ref{likelihood}). Trajectories that satisfy Newton's equations are local extrema of the likelihoods. Again, the development is for nonrelativistic classical particle instances (\ref{L0-ineq}) in a reference frame collocated with the classical center of mass. And, from (\ref{lambda-ineq}), since $L_0(\lambda)$ increases without bound, the accuracy of the nonrelativistic approximation persists.

From the evaluation of the connected contribution (\ref{reslt-connctd}) to the scalar product (\ref{scalarprod}), (\ref{deltat-defn}) of Lemma 2, and the growth (\ref{growth}) of $L_0(\lambda)$,\[C(\lambda,\tau) = {\ds \frac{c_{2n} L_e^3}{(2\pi n)^2 \sqrt{\widehat{{\bf a}^2}}}} \; e^{-n (\widehat{{\bf b}^2}- \widehat{{\bf a}\cdot{\bf b}}^2/ \widehat{{\bf a}^2})/2} e^{-(\delta T)^2/(2n \widehat{{\bf a}^2)}}\]with\[L_e^2 =2L_0(\tau)^2,\qquad \qquad L_s\rightarrow \infty\]and from (\ref{ab-defn}) and (\ref{trans-inv}),\[\widehat{{\bf a}^2} = {\ds \frac{T(\tau)}{nL_0(\tau)^2}},\qquad \qquad \widehat{{\bf a}^2} \widehat{{\bf b}^2} -\widehat{{\bf a}\cdot {\bf b}}^2 = {\ds \frac{I(\tau)T(\tau)-\frac{1}{4} \dot{I}(\tau)^2}{n^2L_0(\tau)^4}}.\]The assertion (\ref{growth}) for the rate of growth of $L_0(\lambda)$ is determined so that $I(\lambda)/L_0(\lambda)^2\rightarrow 0$ and $\lambda^2 T(\lambda)/L_0(\lambda)^2\rightarrow 0$ as $\lambda$ grows without bound. $T(\lambda)$ is bounded and $I(\lambda)$ is asymptotically proportional to $\lambda^2$ when the final states are described by nearly free particles and bound state clusters, the case of interest. The assumed divergence of $L_0(\lambda)$ leads to the significant simplification that $\widehat{{\bf b}^2}- \widehat{{\bf a}\cdot{\bf b}}^2/\widehat{{\bf a}^2}$ is independent of $\lambda$ and the same factor appears in both $C(\lambda,\tau)$ and $C(\tau,\tau)^\frac{1}{2}$.

For sufficiently small $L_0(\tau)$, that is, in point particle-like cases, the connected contribution (\ref{norm-c}) dominates the norm $\|\underline{s}(\tau)\|_{\cal B}$ while the plane wave limit (\ref{norm-wave}) applies for $P_f$. From (\ref{norm-c}),\[\|\underline{s}(\tau)\|_{\cal B}^2\approx c_{2n}C(\tau,\tau)={\ds \frac{L_0(\lambda)^3}{(2\pi n)^2 \sqrt{\widehat{{\bf a}^2}}}} \; e^{-n (\widehat{{\bf b}^2}- \widehat{{\bf a}\cdot{\bf b}}^2/\widehat{{\bf a}^2})/2}\]with\[\widehat{{\bf a}^2} = {\ds \frac{2T(\tau)}{nL_0(\tau)^2}},\qquad \qquad 
\widehat{{\bf b}^2} = {\ds \frac{2I(\tau)}{nL_0(\tau)^2}},\qquad \qquad
\widehat{{\bf a}\cdot {\bf b}} = {\ds \frac{\dot{I}(\tau)}{nL_0(\tau)^2}}\]from (\ref{ab-defn}) and distinct by a factor of two from the evaluation of $\widehat{{\bf a}^2}$, $\widehat{{\bf b}^2}$ and $\widehat{{\bf a}\cdot{\bf b}}$ in $C(\lambda,\tau)$. From the evaluation of the forward contribution (\ref{reslt-forw}) to the scalar product (\ref{scalarprod}), for sufficiently diverse classical trajectories, the forward contribution is negligible with respect to the connected contribution. Then, the indicated scalar product and norms substituted into the expression for the amplitude (\ref{i-defn}) used to evaluate likelihoods (\ref{likelihood}) results in\[\renewcommand{\arraystretch}{2.25} \begin{array}{rl} {\cal I}(\lambda,\tau) &= \sqrt{c_{2n}}\,\left({\ds \frac{\lambda_c}{2}}\right)^\frac{n}{2} {\ds \frac{|C(\lambda,\tau)|}{C(\tau,\tau)^\frac{1}{2}}}\\
 &= \sqrt{c_{2n}}\;k_{\cal I}\; {\ds \frac{L_0(\tau)^2}{T(\tau)^\frac{1}{4}}}\,e^{-L_0(\tau)^2 (\delta T)^2/2T(\tau)}\end{array}\]with $d\mu_s=d({\bf q})_n$, $({\bf q})_n$ are the momenta in the description of $|\underline{s}(\lambda)\rangle$ and\[k_{\cal I} := {\ds \frac{1}{\pi}} \left({\ds \frac{8}{n}}\right)^\frac{3}{4} \left({\ds \frac{\lambda_c}{2}}\right)^\frac{n}{2}.\]A significant simplification is the common factor of\[\exp(-\frac{I(\tau)-\dot{I}(\tau)^2 /T(\tau)}{2L_0(\tau)^2})\]in $C(\lambda,\tau)$ and $C(\tau,\tau)^\frac{1}{2}$. With the designation $T_\infty :=T(\lambda)$ and substitution for $\delta T$ from (\ref{defn-connctd}),\begin{equation}\label{pot-crit} {\cal I}(\lambda,\tau) = \sqrt{c_{2n}}\; k_{\cal I}\; {\ds \frac{L_0(\tau)^2}{T(\tau)^\frac{1}{4}}}\, \exp\left(-\alpha {\ds \frac{L_0(\tau)^2}{T(\tau)}} (T(\tau)-T_\infty)^2\right) \end{equation}with\[\alpha :={\ds \frac{1}{2}\left(\frac{1}{\lambda_c}\right)^2}.\]In this approximation, the likelihood depends only on the initial momentum spread length $L_0(\tau)$, the initial classical kinetic energy $T(\tau)$, and the final classical kinetic energy $T_\infty$. This form indicates that the extrema of (\ref{likelihood}) occur for trajectories $\xi_k(\tau)$ that are the trajectories of classical dynamics.
\newline

\addtocounter{theorems}{1}
{\bf Lemma} \thetheorems: For the VEV given by (\ref{vev-forw}) and (\ref{vev-connctd}) evaluated for minimum packet functions (\ref{n-minpak}) with parameters describing a nonrelativistic classical particle case (\ref{L0-ineq}) and sufficiently small initial $L_0(\tau)$, the trajectories $\xi_k(\tau)$ that maximize the transition likelihoods (\ref{likelihood}) for $\lambda$ increasing without bound and when the momentum spread length $L_0(\lambda)/\lambda \rightarrow \infty$ satisfy Newton's equation (\ref{newton}) with a potential (\ref{potnl}).
\newline

The lemma follows from a demonstration that, in this case, extrema of ${\cal I}(\lambda,0)$ imply extrema of (\ref{likelihood}), and that when the momentum spread length $L_0(\tau)\rightarrow \infty$ sufficiently rapidly and for sufficiently small initial $L_0(\tau)$, the expression (\ref{pot-crit}) for the amplitude ${\cal I}(\lambda,\tau)$ results.

In this case of interest, the projection (\ref{proj}) onto a neighborhood of plane wave states (\ref{norm-wave}) results in the likelihood\[\renewcommand{\arraystretch}{1.75} \begin{array}{rl} \mbox{Trace}(P_f |\underline{u}_s(0)\rangle \langle \underline{u}_s(0)|) &=\left(\frac{\lambda_c}{2}\right)^n {\ds \int_{W_\sigma}}d({\bf q})_n\;{\cal I}(\lambda,0)^2\\
 &\approx {\cal I}(\lambda,0)^2\;\left(\frac{\lambda_c}{2}\right)^n {\ds \int_{W_\sigma}} d({\bf q})_n\end{array}\]from the continuity, indeed infinite differentiability, of the connected contribution (\ref{pot-crit}) in the parameters $({\bf q})_n$ of $\tilde{s}_n(({\bf p})_n;\lambda)$ in the plane wave limit, for sufficiently diverse ${\bf q}_k$ that the forward contribution (\ref{reslt-forw}) is negligible, and for summation over a sufficiently small neighborhood $W_\sigma$. The continuity is with respect to variations in the $n$ parameters $\lambda_c {\bf q}_k = \dot{\xi}_k(\lambda)$ and follows from (\ref{pot-crit}) using the expression for $T_\infty$ as the sum of squares of $\dot{\xi}_k(\lambda)$ from (\ref{it-defn}). As a consequence, the maximization of the ${\cal I}(\lambda,\tau)$ for any $\tau\geq 0$ results in a maximum value for the likelihood (\ref{likelihood}).

From the definition (\ref{it-defn}) of $T(\lambda)$ and the form (\ref{L0-form}) for $L_0(T,V)$, ${\cal I}(\lambda,\tau)$ is a function of the trajectories $\xi_k(\lambda)$ and $\dot{\xi}_k(\lambda)$ evaluated at $\lambda=\tau$ and $\lambda\rightarrow \infty$. The final state consists of plane waves and is considered as fixed. To derive trajectories $\xi_k(\lambda)$ associated with the optimization of the likelihood (\ref{likelihood}), the initial point $\tau$ is varied. Maxima of the likelihood (\ref{likelihood}) occur when the amplitudes ${\cal I}(\lambda,\tau)$ are solutions to the Euler-Lagrange equations. With\[F((\xi(\tau),\dot{\xi}(\tau))_n):= {\cal I}(\lambda,\tau),\]extrema of the amplitude ${\cal I}(\lambda,\tau)$ result when the trajectories $\xi_k(\tau)$ are solutions to\[\frac{\partial F((\alpha,\beta)_n)}{\partial \alpha_k}-\frac{d\;}{d\tau}\left(\frac{\partial F((\alpha,\beta)_n)}{\partial \beta_k}\right)=0\]evaluated with $\alpha_k=\xi_k(\tau)$ and $\beta_k=\dot{\xi}_k(\tau)$. Satisfaction of the Euler-Lagrange equations provides that the first variation $\delta {\cal I}(\lambda,\tau)$ with respect to variations of the trajectories $\xi_k(\tau)$ is zero for any $\tau$. No variation of the trajectories increases summations of the nonnegative likelihood.\[\delta \int_{\tau_a}^{\tau_b} d\tau\;{\cal I}(\lambda,\tau)=0\]when $\xi_k(\tau_a),\dot{\xi}_k(\tau_a)$ and $\xi_k(\tau_b),\dot{\xi}_k(\tau_b)$ are specified.

The form (\ref{L0-form}) for $L_0(\tau)=L_0(T(\tau),V(\tau))$ is selected to result in Newton's equation for the classical trajectories. With ${\cal I}:={\cal I}(\lambda,\tau)$ from (\ref{pot-crit}), the chain rule provides that\[\frac{\partial {\cal I}}{\partial \xi_k}=\frac{\partial V}{\partial \xi_k}\frac{\partial {\cal I}}{\partial V}\]from $\partial T/\partial \xi_k=0$ using (\ref{it-defn}). For potentials that are independent of velocities, $\partial V/\partial \dot{\xi}_k=0$ from (\ref{potnl}) and $\partial T/\partial \dot{\xi}_k=\dot{\xi}_k$ from (\ref{it-defn}). The chain rule provides that\[ \frac{\partial {\cal I}}{\partial \dot{\xi}_k}=\frac{\partial T}{\partial \dot{\xi}_k}\frac{\partial {\cal I}}{\partial T}=\dot{\xi}_k \frac{\partial {\cal I}}{\partial T}.\]The Euler-Lagrange equations for the trajectories that determine the extrema of the likelihood (\ref{likelihood}) are\[\renewcommand{\arraystretch}{2.25} \begin{array}{rl} 0&= {\ds \frac{\partial {\cal I}}{\partial \xi_k}}+ {\ds \frac{d\;}{d\tau} \frac{\partial {\cal I}}{\partial \dot{\xi}_k}}\\
 &={\ds \frac{\partial V}{\partial \xi_k}\frac{\partial {\cal I}}{\partial V}} + {\ds \frac{d\;}{d\tau} \left(\dot{\xi}_k \frac{\partial {\cal I}}{\partial T}\right)}\\
 &= {\ds \frac{\partial V}{\partial \xi_k}\frac{\partial {\cal I}}{\partial V}} +\ddot{\xi}_k {\ds \frac{\partial {\cal I}}{\partial T}} +\dot{\xi}_k {\ds \dot{T} \frac{\partial^2 {\cal I}}{\partial T^2}}+\dot{\xi}_k \dot{V} {\ds \frac{\partial^2 {\cal I}}{\partial V \partial T}}.\end{array}\]Then, should\begin{equation}\label{part-asmpt}\frac{\partial {\cal I}}{\partial T}=\frac{\partial {\cal I}}{\partial V},\end{equation}the Euler-Lagrange equations are satisfied when\[ (\ddot{\xi}_k+\frac{\partial V}{\partial \xi_k}) \frac{\partial {\cal I}}{\partial T} + \dot{\xi}_k (\dot{T}+\dot{V}) {\ds \frac{\partial^2 {\cal I}}{\partial T^2}}=0.\]$L_0(T,V)$ is selected to satisfy (\ref{part-asmpt}) and diverges appropriately for large $\lambda$. If the classical trajectories satisfy Newton's equation,\[\ddot{\xi}_k+\frac{\partial V}{\partial \xi_k}=0,\]then the Euler-Lagrange equations are satisfied since the existence of $V$ implies that the force is conservative [\ref{goldstein}] and it follows that the classical total energy $e_C=T+V$ is a constant of the motion, $\dot{T}+\dot{V}=0$. With selection of the appropriate $L_0(\lambda)$, for this selected UQFT and for these particle-like to wave-like transitions, satisfaction of Newton's equations is a sufficient condition for classical trajectories $\xi_k(\tau)$ to provide extrema of the likelihood (\ref{likelihood}). $L_0(\lambda)$ is constrained to be twice continuously differentiable to apply the Schwarz-Clairaut theorem on the independence from precedence in second partial derivatives. 

This solution for the trajectories also has the property that the maximum of the likelihood (\ref{likelihood}) is nearly independent of the temporal parameter $\tau$ of the initial state.\[\frac{d {\cal I}(\lambda,\tau)}{d\tau}\approx 0.\]This results from the chain rule, and the conservation of the classical energy that results for trajectories that satisfy Newton's equation (\ref{newton}).\[ {\ds \frac{d {\cal I}}{d \tau}} = \dot{T}{\ds \frac{\partial {\cal I}}{\partial T}}+ \dot{V} {\ds \frac{\partial {\cal I}}{\partial V}}\]and the assumed equality of partial derivatives (\ref{part-asmpt}) provides that\[ {\ds \frac{d {\cal I}}{d \tau}} = (\dot{T}+ \dot{V}) {\ds \frac{\partial {\cal I}}{\partial T}}=0\]from $\dot{T}+\dot{V}=0$. The approximation results from the approximations in the evaluation of the amplitude ${\cal I}$ (\ref{pot-crit}).

The equality of partial derivatives (\ref{part-asmpt}) applies if $L_0(T,V)$ satisfies a partial differential equation. From the evaluation (\ref{pot-crit}) of the amplitude ${\cal I}$,\[\renewcommand{\arraystretch}{2.25} \begin{array}{rl} {\ds \frac{\partial {\cal I}}{\partial T}} &=\left(c_0(T,L_0)+d_0(T,L_0) {\ds \frac{\partial L_0}{\partial T}} \right) {\cal I}\\
 {\ds \frac{\partial {\cal I}}{\partial V}} &=d_0(T,L_0) {\ds \frac{\partial L_0}{\partial V}} \,{\cal I}\end{array}\]with\begin{equation}\label{e-l-defn}\renewcommand{\arraystretch}{2.25} \begin{array}{rl} c_0(T,L_0) &:= -{\ds \frac{1}{4T}}-\alpha L_0^2\,\left(1-{\ds \frac{T_\infty^2}{T^2}}\right)\\
 d_0(T,L_0) &:= {\ds \frac{2}{L_0}} -2\alpha L_0\,{\ds \frac{(T-T_\infty)^2}{T}}.\end{array}\end{equation}The selection\[L_0(T,V) = f(T) g(T+V)\]results in an ordinary differential equation for $f(T)$ and $g(s)$ eliminates the trivial solution $\partial {\cal I}/\partial V=\partial {\cal I}/\partial T=0$. (\ref{part-asmpt}) becomes\[c_0(T,fg)+d_0(T,fg) \left( {\ds \frac{\partial f}{\partial T}}\,g +f{\ds \frac{\partial g}{\partial T}}\right) = d_0(T,fg)\, f\,{\ds \frac{\partial g}{\partial V}}\]and the simplification from\[\frac{\partial g(T+V)}{\partial T}=\frac{\partial g(T+V)}{\partial V}\]results in an ordinary differential equation for $f(T)$. Substituting (\ref{e-l-defn}),\begin{equation}\label{diffe-l0}\left(1 -\alpha L_0^2 \frac{(T-T_\infty)^2}{T}\right) \frac{2}{f}\, {\ds \frac{df(T)}{dT}} ={\ds \frac{1}{4T}}+{\ds \frac{\alpha L_0^2}{T^2}} (T^2-T_\infty^2).\end{equation}This differential equation has elementary solutions when either of the terms on the left hand side dominates the other. Dominance by the negative term on the left hand side,\[\alpha L_0^2 \;\frac{(T-T_\infty)^2}{T} \gg 1,\]are the cases of interest to the lemma.

For attractive $1/r$ potentials such as gravity and in scattering instances when particles or bound clusters of particles separate without bound, $V(\tau)\leq V(\lambda)$. Since $T+V$ is a constant of the motion for the trajectories that satisfy Newton's equation, $T(\tau)$ reaches its minimum at $T(\lambda)$. That is, in these cases of interest to the lemma,\[T(\tau)\geq T_\infty\]and the right hand side of (\ref{diffe-l0}) is nonnegative. If the factor multiplying the derivative in (\ref{diffe-l0}) were nonnegative, then $df(T)/dT \geq 0$ and the decline in $T(\tau)$ with increasing $\tau$ implies that $L_0(\tau)$ is declining with time for trajectories that conserve the total classical energy $e_C$. These cases contradict the assumption of the lemma of an increasing $L_0(\tau)$ and require different methods. The cases with negative definite factors multiplying the derivative in (\ref{diffe-l0}) are of interest.

From the definition (\ref{it-defn}) for classical kinetic energy, $0<T(\tau)\ll 1$ in nonrelativistic instances except when the number of particles $n\gg 1$. From the nonrelativistic classical particle conditions (\ref{L0-ineq}),\[\alpha\, L_0(\tau)^2 =\frac{1}{2} \left(\frac{L_0(\tau)}{\lambda_c}\right)^2 \gg 1.\]Then, in the factor multiplying the derivative in (\ref{diffe-l0}), $T(\tau) < \alpha L_0(\tau)^2 (T(\tau)-T_\infty)^2$, except when $(T(\tau)-T_\infty)^2$ is less than $T(\tau)/(\alpha L_0(\tau)^2) \ll 1$. From the conditions of the lemma, $\tau$ is not nearly equal to $\lambda$ and $(T(\tau)-T_\infty)^2$ is small only when there is little change in the classical kinetic energy in the descriptions of initial and final states. Trajectories with sufficiently small $|T(\tau)-T_\infty|$ that escape to unbounded distances imply weak interaction, either great initial separations or a weak potential.

The example of two bodies provides explicit instances. For two bodies, the assumptions of the lemma are that $r\rightarrow \infty$ as $\lambda\rightarrow \infty$. The two body example is described by $L_0(\tau)$, $T_\infty$, the particle mass $m$, the potential strength $g$, and the initial velocity. Specification of the initial velocity is equivalent to specification of the initial potential energy $-g/r(\tau)$ or the initial range $r(\tau)$. With the initial kinetic energy described by the ratio to the final energy, $T(\tau):=(\rho+1)T_\infty$, $\rho>0$ and $|V(\tau)|=\rho T_\infty$, $\dot{r}(\tau)= 2 \sqrt{(1+\rho)T_\infty}$, and $r(\tau)=g/|V(\tau)|$. Then, the instances that violate the conditions of the lemma with positive definite factors multiplying the derivative in (\ref{diffe-l0}), $T(\tau)\geq \alpha L_0(\tau)^2 (T(\tau)-T_\infty)^2$, satisfy\[ (1+\rho) \geq \alpha L_o(\tau)^2 \rho^2 T_\infty.\]This is equivalent to the condition\[ 0< \rho \leq \frac{1+\sqrt{1+4\alpha L_0(\tau)^2 T_\infty}}{2\alpha L_0(\tau)^2T_\infty}\]and a nonnegative coefficient multiplies the derivative in (\ref{diffe-l0}) except for nearly bound instances with more nearly point-like initial descriptions. A large $\rho$ occurs only for small $\alpha L_0(\tau)^2 T_\infty$. Since $\alpha L_0(\tau)^2\gg 1$, a small $\alpha L_0(\tau)^2 T_\infty$ is a $T_\infty$ small with respect to $(\alpha L_0(\tau)^2)^{-1}$. A small $T_\infty$ is a small velocity at escape, that is, a nearly bound instance. For the example of a $10^{-8}$ kg particle mass, $\lambda_c=3.5\times 10^{-35}$ m, and $T_\infty=4.0\times 10^{-6}$ results in $\dot{r}(\lambda)=0.004c$. For $L_0(\tau)=10^4\lambda_c=3.5\times 10^{-31}$ m, $\alpha L_0(\tau)^2=5.0\times 10^7$. Only if the magnitude of the initial potential energy $\rho T_\infty$ decreases below $\rho \leq 0.073$ will the negative definiteness of the coefficient of the derivative in (\ref{diffe-l0}) be violated. However, for smaller initial packet sizes, $L_0(\tau)=10^2 \lambda_c=3.5\times 10^{-33}$ m, $\alpha L_0(\tau)^2=$5000 and $\rho \leq 51$ violate the negative definiteness. For a fixed ratio $L_0(\tau)/\lambda_c$ and a fixed final velocity $\dot{r}(\lambda)$, the value of $\rho$ that violates the assumptions of the lemma is independent of mass $m$.

The nonrelativistic classical particle conditions (\ref{L0-ineq}) must also be satisfied and $r(\tau)\gg \lambda_c$ is violated for small masses or weak interaction. For gravity and two bodies, $-V=\rho T_\infty=Gm/c^2r$ and\[\frac{r(\tau)}{\lambda_c} =\frac{Gm^2}{\hbar c\, \rho T_\infty}\gg 1.\]The analysis of this section does not apply to more point-like initial packet sizes, small masses, and nearly bound or weakly interacting cases. In these cases, the assumptions that result in an $L_0(\tau)$ that increases with time do not apply.

For $T\ll \alpha L_0^2(T-T_\infty)^2$ and trajectories that satisfy Newton's equation, the differential equation (\ref{diffe-l0}) simplifies to\[\frac{df(T)}{dT} \approx \frac{-1}{8\alpha_g (T-T_\infty)^2}\frac{1}{f(T)} -\frac{(T+T_\infty)}{2T(T-T_\infty)} f(T)\]with\[\alpha_g := g(e_C)^2 \alpha\]and has the solution\[f(T)=\left( \frac{T}{4\alpha_g} \ln(\frac{c_L}{T})\right)^\frac{1}{2} \, \frac{1}{T-T_\infty}.\]For trajectories that satisfy Newton's equation, $T+V=e_C$ is a constant of the motion. The definition of $\alpha_g$ results in\begin{equation}\label{L0-reslt} L_0(T,V) =\frac{g(T+V)}{g(e_C)} \left( \frac{T}{4\alpha} \ln(\frac{c_L}{T})\right)^\frac{1}{2} \, \frac{1}{T-T_\infty}.\end{equation}$c_L$ is a constant without units selected to satisfy the constraints (\ref{L0-ineq}) and implement the correspondence of UQFT with a single classical potential, (\ref{lambda-r}). If an $L_0(0)$ is selected to satisfy (\ref{L0-ineq}), then determination of $c_L$ results in\[L_0(\tau)= L_0(0) \left(\frac{T(0)-T_\infty}{T(\tau)-T_\infty}\right)\left(\frac{T(\tau)}{T(0)}\right)^\frac{1}{2} \left(1+\frac{T(\tau)\, \ln(T(0)/T(\tau))}{4\alpha L_0(0)^2 (T(0)-T_\infty)^2}\right)^\frac{1}{2}\]in the abbreviated notation $L_0(\tau)=L_0(T(\tau),V(\tau))$.

Qualitatively from the differential equation (\ref{diffe-l0}) and explicitly from (\ref{L0-reslt}), as time increases, $L_0(\tau)$ increases from an initial value $L_0(0)$ as $T(\tau)$ declines toward $T_\infty$. Eventually, $T(\tau)\approx T_\infty$ and the derivative diverges. However, the assumptions that led to the approximation (\ref{pot-crit}) of the likelihood will be violated before this divergence occurs. The rate of growth of $L_0$ corresponds to the rate of decay of the potential. From conservation of energy for a potential that decays at large separations, $T-T_\infty=V$ and then a potential that decays for large $\lambda$ as $1/r^{1+\epsilon}$ with $\epsilon>0$ results in $L_0/\lambda \sim r^{1+\epsilon}/\lambda \rightarrow \infty$ as $\lambda$ grows without bound. Satisfaction of the desired growth (\ref{growth}) of $L_0$ results from the linear motion, $r \sim \dot{r} \lambda$, of particles or bound state clusters at large $\lambda$. The quantum correction to a $1/r$ potential at large $r$ from Section \ref{sec-twop}, a regularization of the $1/r$ potential, is decisive to Lemma 4. This development verifies that there are cases of interest consistent with an $L_0(\lambda)$ that increases without bound as $\lambda$ grows. 

With appropriate selection of a dynamic model for the momentum spread lengths $L_0(\lambda)$ and for sufficiently energetic scattering events, Newton's equations appear naturally in this optimization of the likelihoods (\ref{likelihood}). Trajectories that satisfy Newton's equations provide the trajectories followed by dominant regions in the support of states as the states evolve in time. Validity of the coplanar propagation bound (\ref{near-cop}) would provide that the trajectories of $1/r$ potentials are distinguished by the greatest upper bounds on likelihoods. Lemma 4 applies for nonrelativistic classical particle approximations (\ref{L0-ineq}) to the states of a UQFT in coordinates collocated with the classical center of mass, excluding instances with more point-like initial packet sizes, small masses, and nearly bound or weakly interacting instances.

\section{Discussion}\label{sec-disc}

Using the correspondence studied by Schr\"{o}dinger and Ehrenfest, this study sought tractable examples to indicate whether classical mechanics provides nonrelativistic approximations to the relativistic quantum physics described by scalar UQFT. Examples demonstrate associations of UQFT with appropriate classical limits. The time translations of the family of minimum packet functions that label states of a UQFT associate with nonrelativistic classical particle dynamics at the level of approximation provided by Newton's equations. The methods of UQFT with the physical interpretation due to Schr\"{o}dinger and Ehrenfest provide a rich alternative to canonical quantization but in an explicitly quantum mechanical setting. The classical limits are correspondences of features in the quantum mechanical description with classical dynamical quantities. These features are evident in an appropriate range of the quantum mechanical descriptions, that is, quantum mechanics is the more general theory and classical mechanics approximates special cases. The explicitly quantum mechanical UQFT are approximated by both Feynman series [\ref{feymns}] and classical limits.

A significant revision to ordinary quantum mechanics in relativistic quantum physics is that critical observables no longer correspond to self-adjoint Hilbert space operators. Significantly, and due to Lorentz invariance, position, implemented as multiplication of the functions that label states by the value of an argument ${\bf x}$, does not define a Hermitian Hilbert space operator [\ref{wigner}] and the associated orthogonal projections onto position eigenstates are inconsistent with causality [\ref{yngvason-lqp}]. Such a correspondence is not general in ordinary quantum mechanics, products of self-adjoint operators such as $X^\frac{3}{2} P X^\frac{3}{2}$ are not Hermitian [\ref{bogo}], but location is not excluded in nonrelativistic physics. If the possibility that multiplication by a field does not necessarily define a Hermitian Hilbert space field operator is considered, then realizations of relativistic quantum physics that exhibit interaction are available [\ref{intro},\ref{gej05}]. The technical conjecture of the canonical formalism, that classical dynamic quantities correspond to Hermitian Hilbert space operators with particular algebraic properties, has known qualifications [\ref{pct},\ref{bogo},\ref{wightman-hilbert}] and precludes alternative constructions for relativistic quantum physics. Poincar\'{e} covariance and causality with interaction in a Hilbert space realization of states with positive energy is tractable when the field is not constrained to be Hermitian and the generator of time translations is allowed to assume consistent forms. The development suggests that in the case of relativistic quantum physics, the appropriate descriptions of quantum dynamics are uniquely quantum mechanical, not elevations of classical descriptions. UQFT achieves Hilbert space realization of relativistic quantum physics but with less familiar descriptions for dynamics.


\section*{References}
\begin{enumerate}
\item \label{schrodinger} E. Schr\"{o}dinger, ``Der stetige \"{U}bergang von der Mikro-zur Makromechanik'', {\em Die Naturwissenschaften}, Vol.~14. Issue~28, 1926, p.~664.
\item \label{ehrenfest} P. Ehrenfest, ``Bemerkung \"{u}ber die angen\"{a}herte G\"{u}ltigkeit der klassichen Machanik
innerhalb der Quanatenmechanik,'' {\em Z. Physik}, Vol.~45, 1927, p.~455.
\item \label{messiah} A. Messiah, {\em Quantum Mechanics, Vol. I}, New York, NY: John Wiley and Sons, 1968.
\item \label{weinberg} S.~Weinberg, {\em The Quantum Theory of Fields, Volume I, Foundations}, New York, NY: Cambridge University Press, 1995. 
\item \label{dirac} P.A.M.~Dirac, {\em The Principles of Quantum Mechanics, Fourth Edition}, Oxford: Clarendon Press, 1958.
\item \label{wigner} T.D.~Newton and E.P.~Wigner, ``Localized States for Elementary Systems'', {\em Rev.~Modern Phys.}, Vol.~21, 1949, p.~400.
\item \label{johnson} G.E.~Johnson, ``Measurement and self-adjoint operators'', May 2014, arXiv:quant-ph/\-1405.\-7224.
\item \label{yngvason-lqp} J.~Yngvason, ``Localization and Entanglement in Relativistic Quantum Physics'', Jan.~2014, arXiv:quant-ph/1401.2652.
\item \label{intro} G.E.~Johnson, ``Introduction to quantum field theory exhibiting interaction'', Feb. 2015, arXiv:math-ph/\-1502.\-07727.
\item \label{gej05} G.E.~Johnson, ``Algebras without Involution and Quantum Field Theories'', March 2012, arXiv:math-ph/1203.2705.
\item \label{feymns} G.E.~Johnson, ``Fields and Quantum Mechanics'', Dec.~2013, arXiv:math-ph/\-1312.\-2608.
\item \label{pct} R.F.~Streater and A.S.~Wightman, {\em PCT, Spin and Statistics, and All That}, Reading, MA: W.A.~Benjamin, 1964.
\item \label{bogo} N.N.~Bogolubov, A.A.~Logunov, and I.T.~Todorov, {\em Introduction to Axiomatic Quantum Field Theory}, trans.~by Stephen Fulling and Ludmilla Popova, Reading, MA: W.A.~Benjamin, 1975.
\item \label{wightman-hilbert} A.S.~Wightman,``Hilbert's Sixth Problem: Mathematical Treatment of the Axioms of Physics'', {\em Mathematical Development Arising from Hilbert Problems}, ed.~by F.~E.~Browder,
{\em Symposia in Pure Mathematics 28}, Providence, RI: Amer.~Math.~Soc., 1976, p.~147.
\item \label{wight} A.S.~Wightman, ``Quantum Field Theory in Terms of Vacuum Expectation Values'', {\em Phys.~Rev.}, vol.~101, 1956, p.~860.
\item \label{borchers} H.J.~Borchers, ``On the structure of the algebra of field operators'', {\em Nuovo Cimento}, Vol.~24, 1962, p.~214.
\item \label{type3} J.~Yngvason, ``The Role of Type III Factors on Quantum Field Theory'', {\em Reports on Mathematical Physics}, Vol.~55(1), 2005, p.~135.
\item \label{vonN} J.~von Neumann, {\em Mathematical Foundations of Quantum Mechanics}, Princeton, NJ: Princeton University Press, 1955.
\item \label{wizimirski} Z.~Wizimirski, ``On the Existence of a Field of Operators in the Axiomatic Quantum Field Theory'', {\em Bull. Acad. Polon. Sci., s\'{e}r. Math., Astr. et Phys.}, Vol.~14, 1966, pg.~91.
\item \label{wight-pt} A.S.~Wightman, ``La th\'{e}orie quantique locale et la th\'{e}orie quantique des champs'', {\em Ann. Inst. Henri Poincar\'{e}}, Vol.~A1, 1964, p.~403.
\item \label{gel2} I.M.~Gel'fand, and G.E.~Shilov, {\em Generalized Functions}, Vol.~2, trans.~M.D.~Friedman, A.~Feinstein, and C.P.~Peltzer, New York, NY: Academic Press, 1968.
\item \label{ruelle} D.~Ruelle, {\em Statistical Mechanics: Rigorous Results}, Reading, MA: W.A.~Benjamin, 1974.
\item \label{rgnewton} Roger~G.~Newton, {\em Scattering Theory of Waves and Particles}, New York, NY: McGraw-Hill, 1966.
\item \label{mp01} G.E.~Johnson, ``Massless Particles in QFT from Algebras without Involution'', May 2012, arXiv:math-ph/1205.4323.
\item \label{limits} G.E.~Johnson, ``Classical limits of unconstrained QFT'', Dec. 2014, arXiv:math-ph/\-1412.\-7506.
\item \label{reed} M. Reed and B. Simon, {\em Methods of Modern Mathematical Physics}, New York, NY: Academic, 1979.
\item \label{goldstein} H.~Goldstein, {\em Classical Mechanics}, Reading, MA: Addison-Wesley, 1950.
\item \label{reeh} H.~Reeh and S.~Schlieder, ``Bemerkungen zur Unit\"{a}r\"{a}quivalenz von Lorentz\-invarianten Feldern'', {\em Nuovo Cimento}, Vol.~22, 1961, p.~1051.
\item \label{segal} I.E.~Segal and R.W.~Goodman, ``Anti-locality of certain Lorentz-invariant operators'', {\em Journal of Mathematics and Mechanics}, Vol.~14, 1965, p.~629.
\item \label{masuda} K.~Masuda, ``Anti-Locality of the One-half Power of Elliptic Differential Operators'', {\em Publ. RIMS, Kyoto Univ.}, Vol.~8, 1972, p.~207.
\end{enumerate}
\end{document}